\documentclass[aps,pra,superscriptaddress,twocolumn,notitlepage,nofootinbib]{revtex4-1}
\usepackage[dvips]{graphicx}
\usepackage{dcolumn}
\usepackage{bm}
\usepackage{amsmath,amsthm,amssymb}
\usepackage{epstopdf}
\usepackage{color}
\usepackage{bbold}
\usepackage{mathtools}
\usepackage[sort&compress]{natbib}
\usepackage{hyperref}
\hypersetup{colorlinks=true,linkcolor=blue,citecolor=blue,urlcolor=black}

\definecolor{azure}{rgb}{0.0, 0.5, 1.0}
\definecolor{amber}{rgb}{1.0, 0.49, 0.0}
\definecolor{red}{rgb}{1.0, 0.1, 0.0}
\definecolor{forestgr}{rgb}{0.13, 0.55, 0.13}

\newcommand{\bra}[1]{\left<#1\right|}
\newcommand{\ket}[1]{\left|#1\right>}
\newcommand{\mbf}{\mathbf}
\newcommand{\Tr}{\mathrm{Tr}}
\renewcommand{\Re}{\mathrm{Re}\,}
\renewcommand{\Im}{\mathrm{Im}\,}

\DeclareMathOperator*{\argmax}{arg\,max}

\newcommand{\D}{\hat{\mathcal{D}}}

\newcommand{\Dv}{\hat{\bm{\mathcal{D}}}}

\begin{document}

\title{Subradiance of multilevel fermionic atoms in arrays with filling \texorpdfstring{$n\geq2$}{n>=2}}

\author{A. Pi\~neiro Orioli}
\affiliation{JILA, NIST, Department of Physics, University of Colorado, Boulder, CO 80309, USA}
\affiliation{Center for Theory of Quantum Matter, University of Colorado, Boulder, CO 80309, USA}

\author{A. M. Rey}
\affiliation{JILA, NIST, Department of Physics, University of Colorado, Boulder, CO 80309, USA}
\affiliation{Center for Theory of Quantum Matter, University of Colorado, Boulder, CO 80309, USA}

\date{\today}

\begin{abstract}

We investigate the subradiance properties of $n\geq 2$ multilevel fermionic atoms loaded into the lowest motional level of a single trap (e.g.~a single optical lattice site or an optical tweezer). As pointed out in our previous work~\cite{PineiroArxiv1907}, perfectly dark subradiant states emerge from the interplay between fermionic statistics and dipolar interactions. While in Ref.~\cite{PineiroArxiv1907} we focused on the $n=2$ case, here we provide an in-depth analysis of the single-site dark states for generic filling $n$, and show a tight connection between generic dark states and total angular momentum eigenstates. We show how the latter can also be used to understand the full eigenstate structure of the single-site problem, which we analyze numerically.
Apart from this, we discuss two possible schemes to coherently prepare dark states using either a Raman transition or an external magnetic field to lift the Zeeman degeneracy.
Although the analysis focuses on the single-site problem, we show that multi-site dark states can be trivially constructed in any geometry out of product states of single-site dark states.
Finally, we discuss some possible implementations with alkaline-earth(-like) atoms such as $^{171}$Yb or $^{87}$Sr loaded into optical lattices, where they could be used for potential applications in quantum metrology and quantum information.

\end{abstract}

\maketitle

%%%%%%%%%%%%%					%%%%%%%%%%%%%
%%%%%%%%%%%%%		SECTION		%%%%%%%%%%%%%
%%%%%%%%%%%%%					%%%%%%%%%%%%%

\section{Introduction}

The coupling of a single emitter, e.g.~an excited atom, to the vacuum electromagnetic field leads to single-particle spontaneous decay.
When more atoms are placed close by, the same coupling to the vacuum makes the atoms exchange photons, leading to coherent dipolar interactions and incoherent cooperative decay processes.
These processes give rise to a broad range of collective phenomena such as subradiance, where destructive interference between different single-particle and cooperative decay processes leads to suppressed decay rates, compared to independent emitters.
Due to their long coherence times, subradiant states can be of potential use for quantum metrology, quantum information and integrated photonics.

The first studies of subradiance (and superradiance) date back to Dicke~\cite{DickePR93}, who studied the radiation properties of closely-packed ensembles of two-level atoms. In the limit where the distances are much smaller than the transition wavelength, $r\ll\lambda_0$, Dicke linked the emission properties of the system with the total angular momentum eigenstates of the collective spin of the atoms. For example, for two atoms the symmetric triplet state $(\ket{ge}+\ket{eg})/\sqrt{2}$ is superradiant, whereas the antisymmetric singlet state $(\ket{ge}-\ket{eg})/\sqrt{2}$ is subradiant. When the atoms are close enough to each other, however, quantum statistics becomes important and can give rise to new subradiant (and superradiant) states, as we will see in this work.

In recent years, subradiance has been the focus of much theoretical research.
So far, most works have studied either two-level systems~\cite{ScullyPRL115,RuostekoskiPRL117,RitschSciRep2015,RitschZoubi_IOP2008,AsenjoPRA95,AsenjoPRX2017,AsenjoAlbrechtNJP2019,ChangPRA99,RitschCardonerPRA100,ZollerPRL122,RobicheauxPRA94,JenPRA96,TudelaPRL115,LukinPRL119,LukinPerczelPRA96,ShahmoonPRL118,LesanovskyBuonaiuto2019,LesanovskyPRA97}, or systems with a $J=0\leftrightarrow1$ transition~\cite{EversPRA75,Ritsch1905.01483,LesanovksyNeedham2019,LesanovskyPRL110,LesanovskyBettlesPRA96,YuanArXiv1907}.
Note that in both cases the systems possess a unique ground state.
Moreover, due to the difficulty of the many-body problem, most studies investigate subradiant states where a single or few excitations are shared among all atoms.
This type of subradiant states has been shown to have interesting applications for quantum memories~\cite{AsenjoPRX2017}, atomic clocks~\cite{ChangPRA99}, mirrors~\cite{ShahmoonPRL118}, excitation transport~\cite{RitschCardonerPRA100,LesanovksyNeedham2019}, or to create topological states~\cite{Syzranov2016,LukinPRL119,LukinPerczelPRA96,LesanovskyBettlesPRA96,YuanArXiv1907} or entangled photons~\cite{TudelaPRL115}.

On the experimental front, however, the observation of subradiance has been challenging.
This is evidenced by the relatively few experiments reporting subradiance~\cite{KaiserPRL116,TemnovPRL95,SolanoNatComm2017,ZhouNature2011,DeVoePRL76,EschnerNature2001,HettichScience2002,ZelevinskyNatPhys2015,JulienneTakasuPRL108,WallraffScience2013}, as compared to its superradiant counterpart.
One of the main difficulties lies on the fact that subradiant states are generally hard to prepare and rather sensitive to imperfections. Moreover, typical proposals for subradiance require the atoms to be very close to each other compared to the wavelength of the transition.

In this work, we consider multilevel fermionic atoms in an array with generic $n\geq2$ atoms per site, and thus extend our previous results~\cite{PineiroArxiv1907}.
The multilevel structure, especially the ground-state degeneracy, makes this problem hard to address.
Because of this, only few works have looked into subradiance for multilevel systems with degenerate ground states~\cite{RitschPRL118,Asenjo1906.02204}.
However, these additional degrees of freedom also provide new possibilities for the creation of interesting quantum states of matter.

The key element in our work is the assumption that the $n$ atoms within a single site are prepared in the lowest motional state and that motional excitations are energetically suppressed.
This can be achieved in an optical lattice or in tweezer arrays by increasing the onsite trapping potential.
In this regime, fermion statistics restricts the allowed states of the internal levels of the atoms, and this in turn blocks certain decay channels that would not be blocked for single atoms.
As shown in Ref.~\cite{PineiroArxiv1907} for filling $n=2$, the combination of fermion statistics with the multilevel structure gives rise to a large set of dark states with remarkable features. They are independent of lattice geometry, in particular they do not require subwavelength arrays, they can support up to one excitation per site, and they can be coherently prepared.

\begin{figure}[t!]
\centering
	\includegraphics[width=\columnwidth]{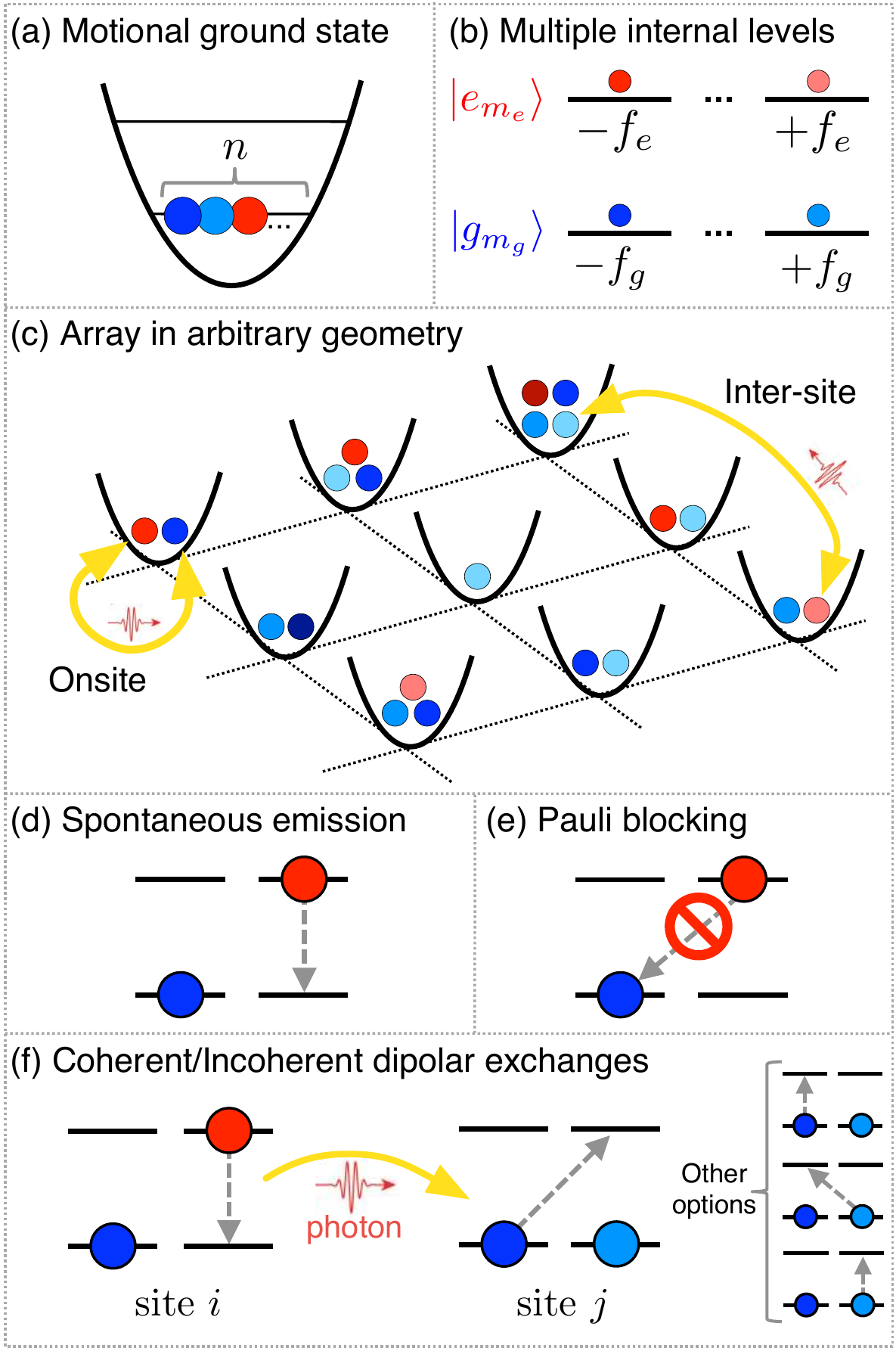}
	\caption{\textbf{Summary of proposed platform.} We consider (a) $n$ fermionic atoms loaded into the motional ground state of a single trap. Each atom has a (b) multilevel internal structure with angular momentum $f_g$ in the ground state (blue colors) and $f_e$ in the excited state (red colors). (c) We consider an arbitrary array of such traps with $n$ atoms per site, which could be realized with an optical lattice or optical tweezers. (d) The atoms can spontaneously decay to different ground states. (e) An atom in a ground state can block the decay of an excited atom due to fermionic statistics. (f) Atoms at different sites, as well as on-site, interact with each other via coherent and incoherent dipolar exchanges mediated by light. Note that the polarizations of the transitions involved need not be the same.}
	\label{fig:platform}
\end{figure}

Here, we extend the study to include higher fillings $n\geq2$ and provide an in-depth analysis of the single-site problem for $n$ dipolar-interacting multilevel fermions in the same motional state. In particular, we show how the basis of total angular momentum states naturally emerges as a suitable description for the eigenstates of this multilevel problem.
Using this, we discuss how dark states can be found in a straightforward manner by looking at the decay channels and multiplicities of the states allowed by fermion statistics.
In addition, we discuss two different schemes to coherently prepare the dark states, using either a Raman-like transition or an external magnetic field, respectively.
The dark states proposed here can be readily implemented e.g.~with alkaline-earth(-like) atoms such as $^{171}$Yb and $^{87}$Sr in optical lattices, and can have potential applications in metrology and quantum information.

This work is structured as follows. In Sec.~\ref{sec:dipolar_master_eq} we introduce the multilevel dipolar master equation used throughout. In Sec.~\ref{sec:totalF} we define and discuss the properties of total angular momentum eigenstates. An in-depth analysis of the dark states is presented in Sec.~\ref{sec:darks}, whereas Sec.~\ref{sec:eigenstates} numerically analyzes the properties of the remaining eigenstates. The two different excitation schemes are detailed in Sec.~\ref{sec:preparation}, possible implementations are discussed in Sec.~\ref{sec:implementation} and concluding remarks are given in Sec.~\ref{sec:conclusions}. The appendices contain additional information on the derivation of the master equation and onsite interaction (Apps.~\ref{app:master_eq_deriv} and \ref{app:onsite}), the properties of the dipole operator (App.~\ref{app:dipole_op}), and the dark states (Apps.~\ref{app:darks_single}, \ref{app:totalFstates} and \ref{app:dark_superposition}).

%%%%%%%%%%%%%					%%%%%%%%%%%%%
%%%%%%%%%%%%%		SECTION		%%%%%%%%%%%%%
%%%%%%%%%%%%%					%%%%%%%%%%%%%

\section{Dipolar master equation\label{sec:dipolar_master_eq}}

We consider arrays of multilevel fermionic atoms with $n$ atoms per site, which can be realized by loading the atoms into, e.g., an optical lattice~\cite{GobanNature2018} or tweezers.
We assume that the atoms are pinned at their corresponding site and that they are in the motional ground-state of the trapping potential.
This can be achieved in the limit of a deep trapping potential, such that tunneling between different sites is suppressed, and the energy to the next motional band (which is of the order of the onsite trapping frequency) is much larger than typical atom-atom interactions and the lattice photon recoil energy.
The latter is characterized by the Lamb-Dicke parameter $\eta\ll1$ with $\eta=2\pi \ell/\lambda_L$~\cite{ZollerSandnerPRA2011}, where $\ell=\sqrt{\hbar/(2m\omega)}$ is the width of the ground-state wave function, $\lambda_L$ is the wave-length of the trapping laser, and $\omega$ is the trapping frequency.

Each atom is a multilevel fermion with an internal level structure consisting of a ground state manifold with total angular momentum $f_g$ and an excited state manifold with $f_e$, separated by an energy $\omega_0=ck_0$.
We consider three different types of level structures relevant for one-photon transitions:
\begin{align}
\begin{aligned}
	\text{multi-}\Lambda: &\, \quad f_e=f_g-1,\\
	\text{multi-}\square: &\, \quad f_e=f_g,\\
	\text{multi-}V: &\, \quad f_e=f_g+1.
\end{aligned}
\label{eq:level_structures}
\end{align}
We label the set of $2f_g+1$ degenerate ground states by $\ket{g_{m_g}}\equiv\ket{g,f_g,m_{g}}$ with $m_g\in[-f_g,f_g]$, and the set of $2f_e+1$ degenerate excited states by $\ket{e_{m_e}}\equiv\ket{e,f_e,m_{e}}$ with $m_e\in[-f_e,f_e]$.
Unless stated otherwise, the ground (excited) states will be assumed to be degenerate.

%%% ----------------------------------------------------------------------------
%%%				Subsection
%%% ----------------------------------------------------------------------------

\subsection{Light-matter Hamiltonian\label{ssec:light_matterH}}

We consider the coupling of the atoms to the electromagnetic field in dipole approximation.
The total light-matter Hamiltonian for this system is given by
\begin{equation}
	\hat{H}_\text{tot} = \hat{H}_\text{atom} + \hat{H}_\text{field} + \hat{H}_\text{af}.
\end{equation}
Since the atoms are assumed to occupy the motional ground state only, we do not include here the motional energies.
The second-quantization Hamiltonian for the atomic energies is given by
\begin{equation}
	\hat{H}_\text{atom} = \hbar\omega_0 \sum_{m} \int d\mbf r\, \hat\sigma_{e_me_m}(\mbf r)
\label{eq:Hatom}
\end{equation}
Here, $\hat{\sigma}_{a_m b_n}(\mbf r)\equiv \hat\psi^\dagger_{a_m}(\mbf r) \hat\psi_{b_n}(\mbf r)$ and $\hat\psi^{(\dagger)}_{a_m}(\mbf r)$ are annihilation (creation) operators of fermions in the internal level $|a_m\rangle$, with $a,b\in\{g,e\}$. They fulfill the anticommutation relations $\{ \hat\psi_{a_m}(\mbf r) , \hat\psi^\dagger_{b_n}(\mbf r') \} = \delta_{ab}\,\delta_{mn}\, \delta(\mbf r-\mbf r')$.

For the light field we consider a discrete set of vacuum modes in a finite volume $V$\footnote{The limit $V\rightarrow\infty$ is taken at the end.} given by
\begin{equation}
	\hat{H}_{\text{field}} = \sum_{\mbf k,\lambda} \hbar\omega_k \left( \hat{a}^\dagger_{\mbf k,\lambda} \hat{a}_{\mbf k,\lambda} + \frac{1}{2} \right),
\label{eq:Hfield}
\end{equation}
where $\omega_k=ck$ and the electromagnetic vacuum modes fulfill $ \left[ \hat{a}_{\mbf k,\lambda} , \hat{a}^\dagger_{\mbf k',\lambda'} \right] = \delta_{\mbf k \mbf k'} \delta_{\lambda \lambda'}$, with $\lambda$ labeling the two polarization modes.

The atom-field coupling in the dipole approximation is given by $ \hat{H}_{\text{af}} = - \hat{\mbf d} \cdot \hat{\mbf E}$\footnote{This is valid in the limit $\mbf k\cdot\mbf r \ll1$, where $\mbf k$ is the wave-vector of the electromagnetic mode and $\mbf r$ gives the typical spatial extension of the atom.}
with dipole operator $\hat{\mbf d}=e\mbf r$ and electric field operator $\hat{\mbf E}$.
This can be expanded as
\begin{align}
	\hat{H}_\text{af} =&\, - \sum_{m,n} \sum_{\mbf k,\lambda} \int d\mbf r\, \mathrm{g}_k \Big( \mbf d_{e_m g_n}\, \hat\sigma_{e_mg_n}(\mbf r) + \text{h.c.} \Big) \nonumber\\
	&\,\qquad\qquad\qquad\qquad \cdot\Big( \boldsymbol\epsilon_{\mbf k,\lambda} \hat{a}_{\mbf k,\lambda} e^{i\mbf k\mbf r} + \text{h.c.} \Big),
\label{eq:Hatomfield}
\end{align}
where $\mbf d_{e_m g_n} \equiv \bra{e_m} \hat{\mbf d} \ket{g_n}$ is the dipole matrix element between the atomic states $e_m$ and $g_n$, and $\mathrm{g}_k\equiv\sqrt{\hbar\omega_k/(2\varepsilon_0V)}$ with dielectric constant $\varepsilon_0$.

The matrix elements of the dipole operator can be expanded using the Wigner-Eckart theorem~\cite{brown2003rotational} as $\mbf d_{e_m g_n} \equiv \mbf d^\text{sph}_{mn}\,d^\text{rad}_{ge}/\sqrt{2f_e+1}$, where $\mbf d^\text{sph}_{mn}$ is the spherical and $d^\text{rad}_{ge}$ the radial part.
The radial part can be related to the total decay rate $\Gamma\equiv\sum_n \Gamma_{e_m\rightarrow g_n}$ by
\begin{equation}
	\Gamma = \frac{\omega_0^3}{3\pi\epsilon_0\hbar c^3} \, \frac{|d^{\text{rad}}_{ge}|^2}{2f_e+1}.
\label{eq:Gamma_radialdipole}
\end{equation}
The spherical part can be expanded in Clebsch-Gordan coefficients as ($q=0,\pm1$) (see App.~\ref{app:dipole_op})
\begin{equation}
	\mbf d^\text{sph}_{m+q,m} = C^q_m\, \mbf e_{q}^*,
\label{eq:dsph_clebsch}
\end{equation}
where $C^q_m\equiv \langle f_g,m;1,q | f_e,m+q \rangle$.
The polarization vectors are defined as $\mbf e_0\equiv\mbf e_z$ and $\mbf e_\pm\equiv\mp(\mbf e_x\pm i\mbf e_y)$, where $\mbf e_z$ defines the quantization axis.
Note that the coefficients $C^q_m$ depend on $f_g$ and $f_e$, but we suppress it in the notation for simplicity.

%%% ----------------------------------------------------------------------------
%%%				Subsection
%%% ----------------------------------------------------------------------------

\subsection{Master equation}

Expanding the atomic field operators in the Wannier basis of the lattice, and using a standard Born-Markov approximation to trace out the photon degrees of freedom~\cite{GrossHarochePRep1982,JamesPRA47,LehmbergPRA2} (see App.~\ref{app:master_eq_deriv}), we derive a master equation $d\hat\rho/dt = -i \big[ \hat{H} , \hat\rho \big] + \mathcal{L}(\hat\rho)$ for the density matrix $\hat\rho$ of the atomic excitations with
\begin{align}
	\hat{H} =&\, - \sum_{i,j} \sum_{q,q'} \mathcal{R}_{q,q'}^{i,j}\, \D^+_{i,q}\, \D^-_{j,q'} ,
\label{eq:HLdipoles}\\
	\mathcal{L}(\hat{\rho}) =&\, - \sum_{i,j} \sum_{q,q'} \mathcal{I}_{q,q'}^{i,j} \left( \left\{ \D^+_{i,q}\, \D^-_{j,q'}, \hat{\rho} \right\} - 2 \D^-_{j,q'}\, \hat{\rho}\, \D^+_{i,q} \right).\nonumber
\end{align}
The Hamiltonian part accounts for coherent dipolar exchange interactions between the atoms (collective Lamb shifts), whereas the Linbladian part accounts for incoherent exchange and contains both single-particle spontaneous decay as well as cooperative decay.

The $\D^\pm_{i,q}$ operators can be seen as multilevel raising and lowering operators.
These operators essentially correspond to the spherical part of the dipole operator, and the three components ($q=0,\pm1$) form a vector (see App.~\ref{app:dipole_op}). They are defined as $\hat{\mathcal{D}}^+_{i,q}\equiv(\hat{\mathcal{D}}^-_{i,q})^\dagger$ and
\begin{equation}
	\D^-_{i,q} = \sum_{m} C^q_m\, \hat{\sigma}^{(i)}_{g_m e_{m+q}}.
\label{eq:D-def}
\end{equation}
Here, $\hat \sigma^{(i)}_{g_me_n}\equiv \hat c^\dagger_{i,g_{m}}\hat c_{i,e_{n}}$,
where $\hat c^{(\dagger)}_{i,a_m}$ annihilates (creates) a fermion at the lowest band of lattice site $i$ with internal level $\ket{a_m}$ ($a=g,e$), and $\{ \hat c_{i,a_m} , \hat c^\dagger_{j,b_n} \} = \delta_{ij}\,\delta_{ab}\,\delta_{mn}$.
In essence, the operator $\D^-_{i,q}$ is a sum over all possible decay processes of any of the atoms at site $i$ from an excited level $\ket{e_{m+q}}$ to $\ket{g_m}$ for fixed polarization $q$ and weighted by the Clebsch-Gordan coefficient $C^q_m$.
Thus, the master equation describes all photon-mediated exchange processes between two atoms where an excited atom at site $j$ decays to a ground state via polarization $q$, and a ground atom at site $i$ is excited to an excited state via polarization $q'$.

The strength and viability of the exchange process depends on the polarizations $q$ and $q'$ of the involved transitions and on the relative distance between the atoms through the coefficients
\begin{align}
	\mathcal{R}_{q,q'}^{i,j}\equiv&\, \left( \mbf e_q^{*T}\, \Re G^{ij}\, \mbf e_{q'} \right), \\
	\mathcal{I}_{q,q'}^{i,j}\equiv&\,  \left( \mbf e_q^{*T}\, \Im G^{ij}\, \mbf e_{q'} \right).
\end{align}
For atoms at different sites ($i\neq j$), these coefficients can be written as $G^{ij} = G(\mbf r_i-\mbf r_j)$, where $G$ is (proportional to) the electromagnetic dyadic Green's tensor in vacuum~\cite{novotny2006principles}
\begin{align}
	G(\mbf r) =&\, \frac{3\Gamma}{4} \left\{ \big[ \mathbb{1} - \hat{\mbf r}\otimes \hat{\mbf r} \big] \frac{e^{ik_0r}}{k_0r} \right. \nonumber\\
	&\,\quad \left.+ \big[ \mathbb{1} - 3\, \hat{\mbf r}\otimes \hat{\mbf r} \big] \left( \frac{ie^{ik_0r}}{(k_0r)^2} - \frac{e^{ik_0r}}{(k_0r)^3} \right) \right\},
\label{eq:Gtensor}
\end{align}
where $\hat{\mbf{r}}\equiv \mbf r/|\mbf r|$.

The onsite ($i=j$) interaction coefficients are given by an integral of the dyadic Green's tensor over the Wannier functions (see App.~\ref{app:onsite}) and generally depend on the geometry of the onsite trapping potential. While our main results are independent of the specific shape of the trap, we will discuss some qualitative consequences of the onsite terms on the eigenstates of the system.
For that we will consider the limit of a deep axially-symmetric onsite trapping potential ($k_0r\ll1$), and define the axial symmetry axis as $\mbf e_z^L$. Notice that $\mbf e_z^L$ need not be parallel to the quantization axis $\mbf e_z$. The onsite coefficients can then be written as
\begin{align}
	\Re G^{ii} \approx&\,  \frac{3\Gamma}{4} U(k_0,\boldsymbol\ell) \left( \mathbb{1}-3\,\mbf{e}^L_z \otimes \mbf{e}^L_z \right) ,
\label{eq:onsite_ReG}\\
	\Im G^{ii} \approx&\, \frac{\Gamma}{2} \mathbb{1},
\label{eq:onsite_ImG}
\end{align}
where the function $U(k_0,\boldsymbol\ell)$ controls the strength of the coherent interactions and depends on the wave-function widths $\boldsymbol\ell=(\ell_x,\ell_y,\ell_z)$ with $\ell_x=\ell_y\equiv \ell_\perp$.
Specifically, it is parametrically given by  $U(k_0,\boldsymbol\ell) \propto ( \lambda_0/\lambda_L )^3\, (\nu_z\nu_\perp)^{3/8}$, where $\nu_z$ is the depth of the lattice potential along the $\mbf{e}^L_z$ direction in units of the recoil energy, and $\nu_\perp$ the depth of the potential in the perpendicular directions.
In the limit of a radially symmetric trap one finds $U(k_0,\boldsymbol\ell)=0$.
A more detailed analysis of this function is provided in App.~\ref{app:onsite}.

%%%%%%%%%%%%%					%%%%%%%%%%%%%
%%%%%%%%%%%%%		SECTION		%%%%%%%%%%%%%
%%%%%%%%%%%%%					%%%%%%%%%%%%%

\section{Total angular momentum basis and dipole operator\label{sec:totalF}}

Throughout this work it will prove useful to use the basis of the total angular momentum eigenstates to describe the internal state of the atoms. This is naturally expected given that $\D^\pm_{i,q}$ are spherical operators.
For this purpose, we start by defining for a given site a single-particle Fock basis. For filling $n=2$ we have (dropping the index $i$)
\begin{equation}
	\ket{a_m b_n} \equiv \hat{c}^\dagger_{a_m} \hat{c}^\dagger_{b_n} \ket{\text{vacuum}},
\label{eq:Fockstates_def}
\end{equation}
where $a,b\in\{g,e\}$ and the two atoms are assumed to occupy the same motional ground-state of the lattice site. Notice that $\ket{a_m b_n}=-\ket{b_n a_m}$ due to fermion statistics.
Fock states for higher fillings, $n\geq3$, are similarly defined as $|a_{m} b_{n} c_{p}\ldots \rangle \equiv \hat{c}^\dagger_{a_m} \hat{c}^\dagger_{b_n} \hat{c}^\dagger_{c_p}  \ldots \ket{\text{vacuum}}$.

The total angular momentum of $n$ atoms is defined as $\mbf{F}=\mbf{f}_1+\mbf{f}_2+\ldots+\mbf{f}_n$, where $\mbf{f}_k$ is the total angular momentum of atom $k$, which can be either $f_g$ or $f_e$. For two atoms per site, $n=2$, we define the total angular momentum basis as\footnote{For $a=b$ these states have to be normalized by $1/\sqrt{2}$.}
\begin{align}
	\ket{F,M}_{ab} \equiv \sum_{m_1+m_2=M} \langle f_a,m_1; f_b,m_2 | F, M\rangle \ket{a_{m_1}b_{m_2}},
\label{eq:FM_statesn2_def}
\end{align}
where again $a,b\in\{g,e\}$ and the sum runs over all allowed values of $m_1$ and $m_2$. Here, $\langle f_a,m_1; f_b,m_2 | F, M\rangle$ is a Clebsch-Gordan coefficient.
Due to the properties of addition of angular momenta, the allowed values of $F$ go from $f_a+f_b$ down to $|f_a-f_b|$ in integer steps. However, since we are assuming the atoms to occupy the same motional state (i.e.~the spatial part is symmetric under exchange), fermion exchange symmetry implies that only some of these values are permitted when the two atoms are identical ($a=b$), i.e.~for $\ket{F,M}_{gg}$ and $\ket{F,M}_{ee}$. In this case, only even values of the total angular momentum are allowed ($F=0,2,4,\ldots$) \cite{devanathan2006angular}.
When the atoms are distinguishable ($a\neq b$), all values of $F$ are permitted.

For $n\geq3$ the total angular momentum basis is defined in a similar way, except that the expansion of total angular momentum eigenstates into single-particle Fock states becomes more intricate. Fortunately, for our analysis it will be sufficient to simply know which states are allowed by exchange symmetry.
For $n$ identical particles this can be straightforwardly found using the method of Ref.~\cite{devanathan2006angular} of counting states and their total magnetic number, see~App.~\ref{app:totalFstates}.
When not all atoms are identical, e.g.~when some atoms are in $g$ and others in $e$, we can find the allowed states by first finding the symmetry-allowed states of all the atoms in $g$ ($e$) and then combining the two results using the rules of angular momentum addition, see examples further below and in Apps.~\ref{app:totalFstates} and \ref{app:dark_superposition}.

For the analysis of dark states and the structure of eigenstates, it is important to understand the action of $\D^-_{i,q}$ ($\D^+_{i,q}$) on the atomic states.
In general, the action of $\D^-_{i,q}$ reduces the total number of excitations by one, $n_e\rightarrow n_e-1$, and the total magnetic number by $q$ as $M\rightarrow M-q$.
Here, we defined $n_e$ and $M$ as the eigenvalues with respect to the operators $\hat n_e \equiv \sum_{i,m} \hat c^\dagger_{i,e_m}\hat c_{i,e_m}$ and $\hat M \equiv \sum_{i,m} m\, \hat c^\dagger_{i,e_m}\hat c_{i,e_m} + \sum_{i,n} n\, \hat c^\dagger_{i,g_n}\hat c_{i,g_n}$. Similarly, $\D^+_{i,q}$ increases $n_e$ by one and $M$ by $q$.

On top of this, because $\D^\pm_{i,q}$ are spherical operators, their action on a total angular momentum state can only change the total $F$ by 0 or $\pm1$. An exception is $F=0\rightarrow0$ which is knowingly forbidden. In general, applying $\D^\pm_{i,q}$ on a state $\ket{(\xi)F,M}$ yields a superposition of states $\ket{(\xi')F',M\pm q}$ with $F'\in\{F,F\pm1\}$, where $\xi$ and $\xi'$ stand for the remaining quantum numbers.
The prefactor in front of each state can in principle be computed using the Wigner-Eckart theorem and Clebsch-Gordan coefficients (see App.~\ref{app:dipole_op}). This can be written as
\begin{align}
	\bra{(\xi')F',M'}\D^\pm_{i,q}\ket{(\xi)F,M}&\, \nonumber\\
	= R(F,F',\xi,\xi')\,&\, \Phi(F,F',M,M',q).
\label{eq:totalF_matrixelements}
\end{align}
The important thing to remember is that it is given by a radial part $R$ which depends only on $F$, $F'$, $\xi$, $\xi'$, and a spherical part $\Phi$ independent of $\xi$, $\xi'$.

%%%%%%%%%%%%%					%%%%%%%%%%%%%
%%%%%%%%%%%%%		SECTION		%%%%%%%%%%%%%
%%%%%%%%%%%%%					%%%%%%%%%%%%%

\section{Dark states\label{sec:darks}}

In this section, we explore the existence of dark states for the system described above.
We present a necessary condition for the atoms to be in a dark state which essentially reduces the problem to a single site, along with an analysis of the $n=2$ dark states of multilevel fermions presented in Ref.~\cite{PineiroArxiv1907}.
We then generalize the search for dark states to fillings $n\geq3$, providing some examples of dark states for various level structures, and shortly commenting on the bosonic case. While we generally work in the basis of total angular momentum states, App.~\ref{app:darks_single} provides a complementary analysis of $n=2$ dark states using the single-particle Fock basis.

%%% ----------------------------------------------------------------------------
%%%				Subsection
%%% ----------------------------------------------------------------------------

\subsection{Dark state condition}

A dark state $\ket{D}$ is defined as an eigenstate of the Hamiltonian $\hat{H}$ with zero decay rate, i.e.~$\mathcal{L}(\ket{D}\!\bra{D})=0$.
Because of this, its time evolution simply results in an overall phase factor.
For the setup introduced above, Eq.~(\ref{eq:HLdipoles}), it is straightforward to show that a sufficient condition to fulfill this is given by
\begin{equation}
	\D^-_{i,q} \ket{D} = 0\quad \forall i,q.
\label{eq:dark_condition}
\end{equation}
Since Eq.~(\ref{eq:dark_condition}) has to be fulfilled for each site $i$ independently from the others, one can construct a family of dark states from product states of single-site dark and ground states.
More specifically, let $\ket{D_{\alpha_i}}_i$, $i\in\mathcal{S}_D$, be arbitrary $i$-site dark states fulfilling $\D^-_{i,q} \ket{D_{\alpha_i}}_i = 0$ $\forall q$.
And let $\ket{G_{\beta_j}}_j$, $j\in\mathcal{S}_G$, be arbitrary $j$-site states with all atoms in the ground-state manifold, which are also trivially dark, $\D^-_{j,q} \ket{G_{\beta_j}}_j = 0$ $\forall q$.
Then, any arbitrary product state given by
\begin{equation}
	\ket{D_{\{\alpha_i\},\{\beta_j\}}} \equiv \bigotimes_{i\in\mathcal{S}_D} \ket{D_{\alpha_i}}_i \bigotimes_{j\in\mathcal{S}_G} \ket{G_{\beta_j}}_j
\label{eq:dark_multisite}
\end{equation}
fulfills Eq.~(\ref{eq:dark_condition}) and is hence a dark state of the multisite system.
Notice that these states are robust against imperfect filling, as long as each site fulfills Eq.~(\ref{eq:dark_condition}) separately.

Dark states of the form (\ref{eq:dark_multisite}) have interesting properties, as noted in Ref.~\cite{PineiroArxiv1907}.
First, they are independent of the array geometry, in particular they do not require subwavelength distances between different sites of the array. Second, they can support many excitations, because each array site can be in a dark state harbouring (at least) one excitation.
These properties essentially follow from the fact that sites which are either in a single-site dark or ground state do not interact with each other.
For the same reason, all dark states of the form (\ref{eq:dark_multisite}) are eigenstates of the Hamiltonian in Eq.~(\ref{eq:HLdipoles}) with zero energy shift, $\hat{H}\ket{D}=0$.
This can be very useful for atomic clocks, because it prevents the appearance of undesirable frequency shifts due to dipole interactions, as well as for the creation of entangled states using superpositions of dark states.

For one atom per site there exist no nontrivial solutions to Eq.~(\ref{eq:dark_condition}).
However, as we will see in the upcoming sections, having two (or more) atoms per lattice site opens the possibility for such dark states to exist.
Because the dark states of Eq.~(\ref{eq:dark_multisite}) are product states of single-site dark states, we focus in the following on the case of a single site and drop the index $i$.

%%% ----------------------------------------------------------------------------
%%%				Subsection
%%% ----------------------------------------------------------------------------

\subsection{Interference and fermion statistics\label{ssec:interference}}

Physically, the condition (\ref{eq:dark_condition}) means that all possible ways of the state $\ket{D}$ to decay to a ground level, i.e.~emit a photon, need to be killed. As we will see, this can be achieved either through destructive interference or using fermion statistics.

The decay of an excitation can happen by emitting a photon in either of the three polarizations, $q=0,\pm1$. According to Eq.~(\ref{eq:dark_condition}), all the decays with a fixed polarization $q$ need to be killed independently from the other polarizations, in other words, emission with orthogonal polarizations can not interfere with each other.
Moreover, in our multilevel system, an excited atom can in principle decay to different ground levels (via different polarizations). Interference of different decay processes, however, can only happen if the final state $\ket{\phi_f}$ is the same.
To see this, note that the action of $\D^-_{q}$ on a given state creates a superposition of final states (with one excitation less) each of whose prefactors needs to vanish identically.
This implies that each possible \emph{decay channel} of the state $\ket{D}$ with polarization $q$ and final state $\ket{\phi_f}$ has to be killed independently from the other polarizations and final states. Thus, condition (\ref{eq:dark_condition}) can also be written as (dropping the index $i$)
\begin{equation}
	\bra{\phi_f} \D^-_q\ket{D}=0,\quad \forall q,\ket{\phi_f}.
\label{eq:dark_condition_qf}
\end{equation}
Each possible pair $(q,\ket{\phi_f})$ gives rise to a separate condition to be fulfilled.
Notice that all channels are nevertheless intertwined in a complex fashion since each state has, in general, multiple decay channels, and each decay is weighted by a different Clebsch-Gordan coeffcient.

A direct consequence of this insight is that only states with the same number of excitations $n_e$ and magnetic number $M$ can interfere with each other. This is because only such states can decay to the same final state $\ket{\phi_f}$ via the same polarization $q$. Therefore, dark states can only be superpositions of states with the same $n_e$ and $M$, which can in turn be used to label the dark states.\footnote{Of course, given any two dark states fulfilling (\ref{eq:dark_condition}), any superposition of them is also dark.}
Notice, however, that dark states can in principle be a superposition of states with different total $F$.

%%% ----------------------------------------------------------------------------
%%%				Subsection
%%% ----------------------------------------------------------------------------

\subsection{Dark states for \texorpdfstring{$n=2$}{n=2} fermions per site\label{ssec:darks_n2}}

We present now all the dark states fulfilling (\ref{eq:dark_condition}) for filling $n=2$ and arbitrary multilevel structure ($|f_e-f_g|\leq1$), c.f.~Ref.~\cite{PineiroArxiv1907}.
We start by illustrating the main concepts in the simplest nontrivial example of a dark state, which can be found for the multi-$\square$ structure $f_g=f_e=1/2$.
In this case, there exists one dark state given by the total angular momentum basis state $\ket{D_0}_{\{\frac{1}{2},\frac{1}{2}\}}=\ket{F=0,M=0}_{ge}$. To see why, notice that for two atoms in the ground manifold there exists in this case only one state allowed by fermion statistics, namely $\ket{F=0,M=0}_{gg}$. Since the dipole transition $0\rightarrow0$ is forbidden due to selection rules, this implies that the $\ket{D_0}$ state can not decay to the ground state and is hence dark.

An alternative explanation can be given as well in terms of the single-particle Fock basis.
In this basis, the dark state reads $\ket{D_0}_{\{\frac{1}{2},\frac{1}{2}\}} =\frac{1}{\sqrt{2}} \left( \ket{g_{1/2}\,e_{-1/2}} - \ket{g_{-1/2}\,e_{1/2}} \right)$.
Applying $\D^-_q$ to this state for $q=\pm1$, i.e.~emission of a circularly polarized photon, we see that the decay of each of the Fock states in the superposition is Pauli blocked by the other atom.
For $q=0$, i.e.~emission of a linearly polarized photon, we get $\D^-_1\ket{D_0}_{\{\frac{1}{2},\frac{1}{2}\}} \propto (C^{0}_{1/2}+C^{0}_{-1/2})\ket{g_{-1/2}\,g_{1/2}}$, which vanishes because for this level structure one has $C^{0}_{1/2}=-C^{0}_{-1/2}$.
Hence, the decay from the two states interferes destructively and the state $\ket{D_0}_{\{\frac{1}{2},\frac{1}{2}\}}$ is dark.

\begin{figure}[t!]
\centering
	\includegraphics[width=\columnwidth]{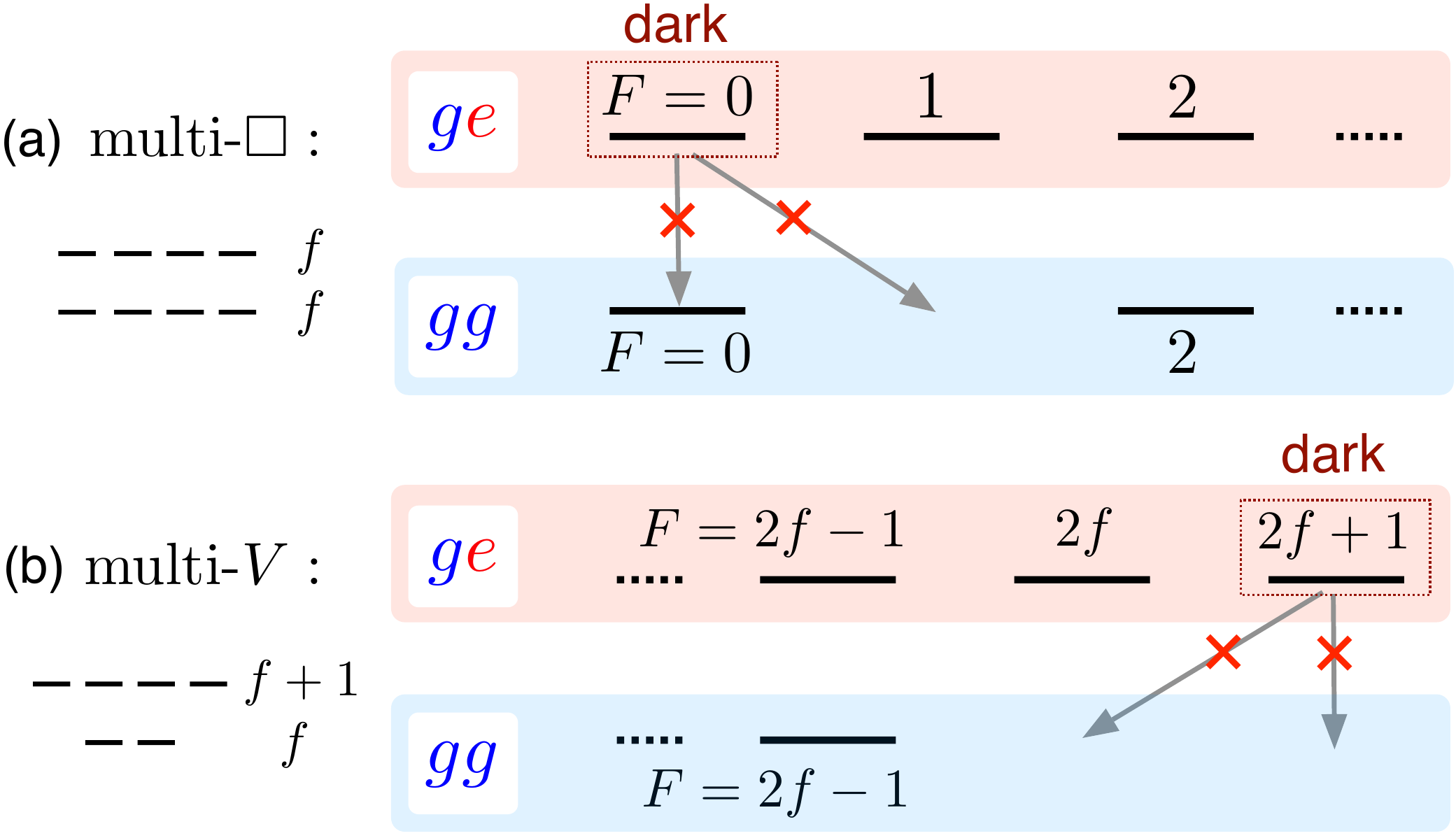}
	\caption{\textbf{Dark states for $n=2$:} Dark states for two fermions on a single site for (a) the multi-$\square$ level structure $(f\leftrightarrow f)$, and (b) the multi-$V$ level structure $(f\leftrightarrow f+1)$. The plot shows a subset of the $gg$ states (no excitations) and the $ge$ states (one excitation) in terms of their total angula momentum. All decay channels of the states $F=0$ and $F=2f+1$ in the multi-$\square$ and multi-$V$ configuration are respectively forbidden.}
	\label{fig:n2darks}
\end{figure}

The dark states can be generalized to generic multi-$\square$ level structures, $f_e=f_g\equiv f$. As we will show in the following sections, for any (half-integer) $f$ there exists exactly one dark state given by (see Fig.~\ref{fig:n2darks})
\begin{align}
	\ket{D_0}_{\{f,f\}} \equiv&\, \ket{F=0,M=0}_{ge}.
\label{eq:dark_n2square}
\end{align}
To understand the possible decay channels of this state, recall first that due to fermion statistics (c.f.~Sec.~\ref{sec:totalF}) only even-numbered total angular momenta (F=0,2,4,\ldots) are allowed when both atoms are in the ground manifold. Since $F$ can only change at most by 1, this means that the above state can only decay to the state $\ket{F=0,M=0}_{gg}$, which again is forbidden due to $0\rightarrow0$. Thus, fermion statistics and dipole selection rules make $\ket{F=0,M=0}_{ge}$ a dark state for the multi-$\square$ level structure.

For generic multi-$V$ level structures, $f_e=f_g+1\equiv f+1$, it can be shown that for each $M\in\{-2f-1,\ldots,2f+1\}$ there exists exactly one dark state given by (see Fig.~\ref{fig:n2darks})
\begin{align}
	\ket{D_M}_{\{f,f+1\}} \equiv \ket{F=2f+1,M}_{ge}.
\label{eq:dark_n2V}
\end{align}
As opposed to the case before, here the dark states are states of maximal total angular momentum. To see why these states are dark, notice first that for both atoms in the ground manifold the maximal total angular momentum allowed by fermion statistics is $F=2f-1$. Hence, due to selection rules ($\Delta F=0,\pm1$) the above state can not decay to any allowed state in the ground-state.
Thus, $\ket{F=2f+1,M}_{ge}$ is a dark state for the multi-$V$ level structure.

For completeness, we note that for the multi-$\Lambda$ structure with $n=2$ atoms per site there exist no dark states, because the number of decay channels is too large. However, for this level structure dark states do exist at larger filling, as we will show below.

As a side remark, we note that these dark states do not generally have any obvious orbital symmetry properties. An exception to this are the dark states $\ket{D_0}_{\{f,f\}}$ and $\ket{D_0}_{\{f,f+1\}}$, which are symmetric and antisymmetric for $g\leftrightarrow e$ exchanges, respectively.\footnote{For comparison, recall that for a two-level system in the Dicke limit the dark state $(\ket{eg}-\ket{ge})/\sqrt{2}$ is antisymmetric. In this case, both atoms are assumed to have the same dipole matrix element connecting their respective $g$ and $e$ states.}
The corresponding symmetry follows from the symmetries of the respective Clebsch-Gordan coefficients in Eq.~(\ref{eq:FM_statesn2_def}) and from $\ket{a_mb_n}=-\ket{b_na_m}$. Specifically, $\langle f_1,m;f_2,-m|F,0\rangle=(-1)^{f_1+f_2-F}\langle f_1,-m;f_2,m|F,0\rangle$.
Alternatively, this can also be understood in the single-particle Fock basis from the symmetries of the $C^q_m$ coefficients.

%%% ----------------------------------------------------------------------------
%%%				Subsection
%%% ----------------------------------------------------------------------------

\subsection{Searching for dark states\label{ssec:dark_search}}

In order to address the more complicated case $n\geq3$, we outline here a general procedure to search for dark states using the basis of total angular momentum states. This will further allow to show that Eqs.~(\ref{eq:dark_n2square}) and (\ref{eq:dark_n2V}) are the only dark states for $n=2$. An alternative proof of this using the single-particle Fock basis is provided in App.~\ref{app:darks_single}.

We divide the search for dark states into three steps:
\begin{enumerate}
	\item Write down all angular momentum states allowed by fermion statistics.
	\item Use dipole selection rules to check if any angular momentum state is dark.
	\item Analyze decay channels of the remaining angular momentum states to search for superposition dark states.
\end{enumerate}
To exemplify this, we focus here on the example of $n=2$ and search for single-excitation dark states.
We start by writing down all states allowed by fermion statistics, both with zero and one excitation.
As argued above (c.f.~Sec.~\ref{sec:totalF}), when both atoms are in the ground manifold only states with even total $F$ are allowed.
For a given half-integer $f_g\equiv f$, this means that the Hilbert subspace of two atoms in the ground manifold is given by
\begin{align}
\begin{aligned}
	&\, \ket{F=0,M}_{gg} \\
	&\, \ket{F=2,M}_{gg} \\
	&\, \ket{F=4,M}_{gg} \\[-5pt]
	&\, \qquad \vdots \\
	&\, \ket{F=2f-1,M}_{gg}
\end{aligned}
\end{align}
where the quantum number $M$ can take any value from $-F$ to $F$, as usual.
States with one $e$ excitation are not restricted by fermion statistics, since the two atoms become distinguishable. The available states now depend on the internal level structure. For the multi-$\square$ structure one has
\begin{align}
\begin{aligned}
	f_e=f_g\equiv f:\quad&\, \ket{F=0,M}_{ge} \\
	&\, \ket{F=1,M}_{ge} \\[-5pt]
	&\, \qquad \vdots \\
	&\, \ket{F=2f,M}_{ge},
\end{aligned}
\label{eq:eg_states_square}
\end{align}
and for the multi-$V$ structure we get
\begin{align}
\begin{aligned}
	f_e=f_g+1\equiv f+1:\quad&\, \ket{F=1,M}_{ge} \\
	&\, \ket{F=2,M}_{ge} \\[-5pt]
	&\, \qquad \vdots \\
	&\, \ket{F=2f+1,M}_{ge}.
\end{aligned}
\label{eq:eg_states_V}
\end{align}
The absence of dark states for the multi-$\Lambda$ level structure can be shown similarly to the absence of other dark states different from Eqs.~(\ref{eq:dark_n2square}) and (\ref{eq:dark_n2V}), and hence will not be further discussed.

Next, we study whether any of the basis states in Eqs.~(\ref{eq:eg_states_square}) and (\ref{eq:eg_states_V}) are dark by looking at their possible decay channels and using dipole selection rules.
It is straightforward to see that all basis states except the ones from Eqs.~(\ref{eq:dark_n2square}) and (\ref{eq:dark_n2V}) have at least one open decay channel.
For example, for the multi-$\square$ structure [Eq.~(\ref{eq:eg_states_square})], the state $\ket{1,M}_{ge}$ can decay to both $\ket{0,0}_{gg}$ as well as $\ket{2,M'}_{gg}$, whereas the state $\ket{2f,M}_{ge}$ can decay to $\ket{2f-1,M'}_{gg}$.

This first analysis, however, does not completely rule out the possibility of dark states made of superpositions of different angular momentum states. In the last step, we consider this possibility.
As shown in Sec.~\ref{ssec:interference}, dark states are made of states with the same magnetic number $M$. Therefore, we start by writing down all available basis states $\ket{F\geq M,M}_{ge}$, omitting the states which have already been shown to be dark. We then analyze all the distinct decay channels with polarization $q$ and final state $\ket{\phi_f}$ of these states. This gives a homogeneous set of equations with $n_\text{decays}$ equations equal to the number of distinct decay channels, and $n_\text{states}$ variables (amplitudes of states) given by the number of available states. Since the Clebsch-Gordan coefficients are real, the variables can be assumed to be real too.

To find out if solutions to the linear system of equations exist, it suffices to know its rank. While this can of course be computed by explicitly writing down all equations, estimating it on general grounds is hard.
For the purpose of estimation, we assume that if all states involved have \emph{different} total $F$, then all equations are linearly independent. Using this, the number of dark states can easily be estimated from $n^M_\text{darks}=n_\text{states}-n_\text{decays}$.
For the cases considered here we find this assumption to be indeed consistent with the numerical results for the total number of dark states, see Sec.~\ref{sec:eigenstates}. In the hypothetical case where this assumption may break, the estimates made here constitute at least a lower bound on the number of dark states for a given system.

Using this one can show that there exist no other dark states for $n=2$ than the ones given in Eqs.~(\ref{eq:dark_n2square}) and (\ref{eq:dark_n2V}). As an example, consider the multi-$\square$ structure with $f>1/2$ and let us look for dark states with $M=2f-1$. In this case, there are two basis states that can be used to form dark states: $\ket{2f,2f-1}_{ge}$ and $\ket{2f-1,2f-1}_{ge}$. Both these states can decay to two states, $\ket{2f-1,2f-1}_{gg}$ and $\ket{2f-1,2f-2}_{gg}$, via polarizations $q=0$ and $q=1$, respectively (recall that $\D^-_q$ leads to $M\rightarrow M-q$). Therefore, the number of dark states is $n^M_\text{darks}=2-2=0$.
The same procedure can be applied to all possible values of $M$ to rule out the possibility of dark states different from Eqs.~(\ref{eq:dark_n2square}) and (\ref{eq:dark_n2V}). Nontrivial superposition dark states will emerge for $n\geq3$, as we show next.

%%% ----------------------------------------------------------------------------
%%%				Subsection
%%% ----------------------------------------------------------------------------

\subsection{Dark states for \texorpdfstring{$n\geq3$}{n>=3} fermions per site\label{ssec:darks_n3}}

Using the concepts introduced in the previous section (Sec.~\ref{ssec:dark_search}), we study the dark states for fillings $n\geq3$. As opposed to the $n=2$ case, finding general analytical expressions for $n\geq3$ dark states for generic level structures is rather complicated. Therefore, we concentrate on some simple cases, derive the total number of dark states, and discuss their general properties.

We start with filling $n=3$ and consider first $f_g=3/2$ since for smaller $f_g$ the states are trivial.
Following the discussion in Sec.~\ref{sec:totalF} only the following states with three atoms in the ground manifold are allowed (see App.~\ref{app:totalFstates}):
\begin{align}
	f_g\equiv f:\quad&\, \ket{F=3/2,M}_{ggg}.
\label{eq:ggg_state}
\end{align}
To write down the allowed single-excitation states $(gge)$ we start by coupling the two atoms in the ground manifold into states $\ket{f_{12},m_{12}}_{gg}$ with $f_{12}=0,2$. This we can straightforwardly couple with the third atom to obtain all the allowed states. To simplify the notation, we will write the allowed states in the format
\begin{align}
	f_g \leftrightarrow f_e:&\,\quad \ket{(f_{12})F,M}_{gge}.
\end{align}
The allowed states for the three level structures of Eq.~(\ref{eq:level_structures}) are then given by
\begin{align}
\begin{aligned}
	\frac{3}{2}\leftrightarrow\frac{1}{2}:\quad&\, \ket{(0)F,M}_{gge}, F=\frac{1}{2}\\
	&\, \ket{(2)F;M}_{gge}, F=\frac{3}{2},\frac{5}{2}
\end{aligned}
\end{align}
\begin{align}
\begin{aligned}
	\frac{3}{2}\leftrightarrow\frac{3}{2}:\quad&\, \ket{(0)F,M}_{gge}, F=\frac{3}{2}\\
	&\, \ket{(2)F;M}_{gge}, F=\frac{1}{2},\frac{3}{2},\frac{5}{2},\frac{7}{2}
\end{aligned}
\label{eq:n3square_states_gge}
\end{align}
\begin{align}
\begin{aligned}
	\frac{3}{2}\leftrightarrow\frac{5}{2}:\quad&\, \ket{(0)F,M}_{gge}, F=\frac{5}{2}\\
	&\, \ket{(2)F;M}_{gge}, F=\frac{1}{2},\frac{3}{2},\frac{5}{2},\frac{7}{2},\frac{9}{2}.
\end{aligned}
\label{eq:n3V_states_gge}
\end{align}
For the multi-$\Lambda$ level structure, $f_g=3/2\leftrightarrow f_e=1/2$ we find again no dark states, because there are too many decay channels.

To find the dark states for the multi-$\square$ and multi-$V$ level structures, it is important to notice that in both cases there exist distinct sets of states with the same total $F$ but different $f_{12}$, see Eqs.~(\ref{eq:n3square_states_gge}) and (\ref{eq:n3V_states_gge}). Because of this, the decay channels of the two sets of states can indeed be shown to be linearly dependent. To see this, consider a superposition $\alpha\ket{a}+\beta\ket{b}$ with the states $\ket{a}\equiv\ket{(f^{a}_{12})F,M}$ and $\ket{b}\equiv\ket{(f^{b}_{12})F,M}$ with $f^a_{12}\neq f^b_{12}$, and assume that they can both decay to $\ket{q}\equiv\ket{F',M-q}$ with $q=0,\pm1$. This gives rise to three equations to be fulfilled, $\bra{q}\D^-_q\ket{a} \alpha + \bra{q}\D^-_q\ket{b} \beta=0$, one for each value of $q$. Using Eq.~(\ref{eq:totalF_matrixelements}) one can show, however, that
\begin{equation}
	\frac{\bra{q}\D^-_q\ket{a}}{\bra{q}\D^-_q\ket{b}} = \frac{R(F,F',f^a_{12})}{R(F,F',f^b_{12})},
\end{equation}
where $R$ is a reduced dipole moment.
This means that the ratio of the prefactors of $\alpha$ and $\beta$ is $q$-independent, and hence that the three equations are exactly equivalent up to an overall prefactor.

For the multi-$\square$ level structure, $f_g=3/2\leftrightarrow f_e=3/2$ we find a total of 12 dark states. First of all, the basis states $\ket{(2)7/2,M}_{gge}$ are automatically dark because the maximal $F$ of the $ggg$ states is 3/2. Apart from this, we can form dark states combining the states $\ket{(0)3/2,M}_{gge}$ and $\ket{(2)3/2,M}_{gge}$. Following the previous argument, these two states (with the same $F$) can decay to $\ket{3/2,M-q}_{ggg}$, where the three conditions $(q=0,\pm1)$ are linearly dependent, i.e.~effectively they have only one decay channel. Thus, there is a superposition which is dark for every $M$. As a proof of principle, we compute this superposition explicitly in App.~\ref{app:dark_superposition}. Therefore, there are a total of $8+4=12$ dark states.

For the multi-$V$ level structure, $f_g=3/2\leftrightarrow f_e=5/2$ we find a total of 24 dark states. In this case, both the $\ket{(2)9/2,M}_{gge}$ and $\ket{(2)7/2,M}_{gge}$ basis states are dark. Apart from this, there exist dark states made up of superpositions of $\ket{(0)5/2,M}_{gge}$ and $\ket{(2)5/2,M}_{gge}$ for each $M$. Thus, the total is $10+8+6=24$.

The same game can be played for level structures with $f_g>3/2$, which leads to ever increasing numbers of dark states. In this way, for example, we obtain 30 dark states for $f_g=5/2\leftrightarrow f_e=5/2$, 60 for $f_g=5/2\leftrightarrow f_e=7/2$, and again none for $f_g=5/2\leftrightarrow f_e=3/2$. Finding general results for dark states made of superpositions of different basis states is complicated for $n=3$. However, we can easily generalize which single basis states are dark for generic level structures. To do so, note that the maximal $F$ of $ggg$ states is $3f_g-3$ (see App.~\ref{app:totalFstates}), whereas the maximal $F$ of $gge$ states is $2f_g-1+f_e$. The difference is $f_e-f_g+2$. Therefore, for the multi-$\square$ structure ($f\leftrightarrow f$) the states $\ket{(2f-1)3f-1,M}_{gge}$ are always dark, whereas for the multi-$V$ structure ($f\leftrightarrow f+1$) the states $\ket{(2f-1)3f,M}_{gge}$ and $\ket{(2f-1)3f-1,M}_{gge}$ are dark.
In neither of the above examples, however, do we find double-excited dark states.

For filling $n=4$ we do find explicit double-excited dark states. As an example we consider $f_g=3/2\leftrightarrow f_e=5/2$. In this case, one can show that there exist 21 single-excitation dark states, specifically the $ggge$ states with $F=2,3,4$. To find the double-excited states recall first that for three atoms in the ground manifold the only allowed state is $\ket{3/2,M}_{ggg}$ [see Eq.~(\ref{eq:ggg_state})]. Therefore, the single-excited states for $n=4$ are given by
\begin{align}
\begin{aligned}
	\frac{3}{2} \leftrightarrow \frac{5}{2}:\quad&\, \ket{F,M}_{ggge}, F=1,2,3,4.
\end{aligned}
\end{align}
To find the available double-excited states we first construct $\ket{f_{12},m_{12}}_{gg}$ with $f_{12}=0,2$, and $\ket{f_{34},m_{34}}_{ee}$ with $f_{34}=0,2,4$. Coupling the two-atom states together then gives rise to the following double-excited states (notation: $\ket{(f_{12},f_{34})F,M}_{ggee}$):
\begin{align}
\begin{aligned}
	\frac{3}{2} \leftrightarrow \frac{5}{2}:\quad&\, \ket{(0,0)F,M}_{ggee}, F=0 \\
	&\, \ket{(0,2)F,M}_{ggee}, F=2 \\
	&\, \ket{(0,4)F,M}_{ggee}, F=4 \\
	&\, \ket{(2,0)F,M}_{ggee}, F=2 \\
	&\, \ket{(2,2)F,M}_{ggee}, F=0,1,2,3,4 \\
	&\, \ket{(2,4)F,M}_{ggee}, F=2,3,4,5,6.
\end{aligned}
\end{align}
From this one can immediately see that the double-excited states $\ket{(2,4)6,M}_{ggee}$ are dark, because the maximal $F$ of the $ggge$ states is 4. Moreover, using the same method of counting the multiplicities of states with the same $F$ and their decay channels (taking their linear dependence into account) reveals that there exist double-excited dark states for superpositions of $F=0,2,4$ states, respectively. This gives a total of $13+1+5+9=28$ double-excited dark states.

For the multi-$\square$ structure with $f_g=3/2\leftrightarrow f_e=3/2$ and $n=4$ we also find one double-excited dark state, made of a superposition of two $F=0$ $ggee$ states. Apart from this, it has 13 single-excited dark states corresponding to $ggge$ states with $F=0,2,3$. For the multi-$\Lambda$ structure with $n=4$ we find no double-excited dark states, but we do find for the first time 5 single-excited dark states with total $F=2$.
This is because while two of the atoms block two ground levels, the remaining two atoms see an effective $1/2\leftrightarrow1/2$ level structure (with Clebsch-Gordan coefficients determined by the original $3/2\leftrightarrow1/2$ transitions).

Dark states for yet higher fillings can be found using the same procedures, or with numerical simulations.

%%% ----------------------------------------------------------------------------
%%%				Subsection
%%% ----------------------------------------------------------------------------

\subsection{Bosonic dark states}

Of course, the formalism introduced here is readily applicable to the bosonic case too, where bosonic statistics can help constructing dark states. For filling $n=2$, the available states with two bosonic atoms in the ground manifold are given by
\begin{align}
\begin{aligned}
	&\,\ket{F=0,M}_{gg} \\
	&\,\ket{F=2,M}_{gg} \\
	&\,\qquad \vdots \\
	&\,\ket{F=2f,M}_{gg},
\end{aligned}
\end{align}
where now $f$ is assumed to be an integer. In this case, for the multi-$\square$ structure $f_e=f_g\equiv f$ the state $\ket{F=0,M=0}_{ge}$ is again a dark state, because it can only decay to $\ket{0,0}_{gg}$ which is again forbidden by selection rules. However, for the multi-$V$ structure there are now no dark states at all. In particular, the states $\ket{F=2f+1,M}_{ge}$ have now an open decay channel to $\ket{2f,M'}_{gg}$, as opposed to the fermionic case.

The special case of effective two-level bosonic atoms is interesting. This effective level structure can be achieved using a strong magnetic field to isolate two stretched levels, without populating the other levels.
In this case, the state $(\ket{ge}-\ket{eg})/\sqrt{2}$ is not allowed by boson statistics when both atoms are in the ground motional state. The only allowed states in that case are the symmetric states, which are not dark. Thus, in order to create dark states with 2 bosonic two-level atoms on a single site, one would need to involve higher motional bands.
A thorough analysis of the dark states of bosons for higher bands or for general filling $n$ is beyond the scope of this work.

%%%%%%%%%%%%%					%%%%%%%%%%%%%
%%%%%%%%%%%%%		SECTION		%%%%%%%%%%%%%
%%%%%%%%%%%%%					%%%%%%%%%%%%%

\section{Numerical eigenstates\label{sec:eigenstates}}

\begin{figure*}[t!]
\centering
	\includegraphics[width=\textwidth]{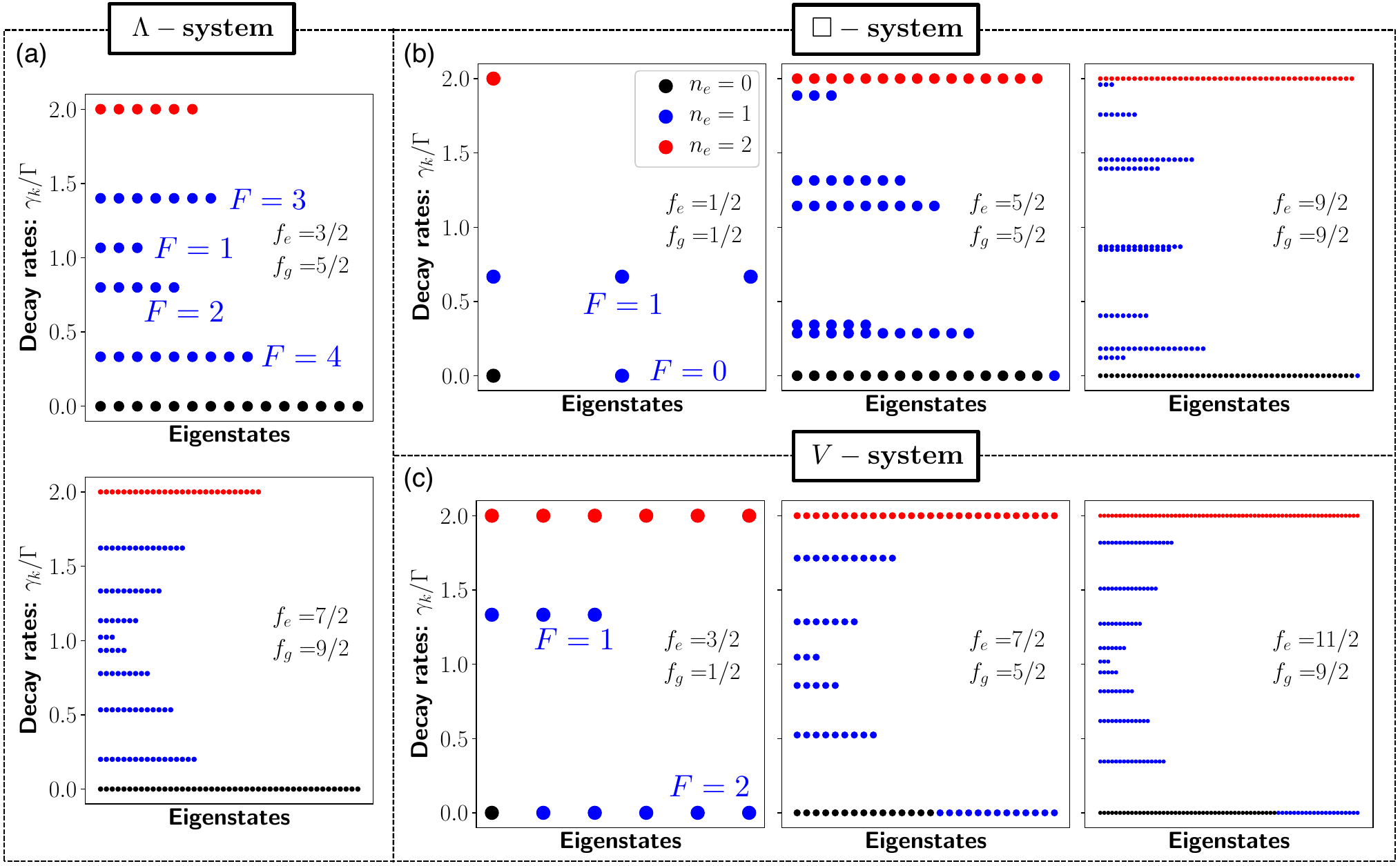}
	\caption{\textbf{Eigenstates for $n=2$.} Decay rates $\gamma_k/\Gamma$ of all eigenstates $\ket{k}$ for different level structures (indicated inside the plots) for $n=2$ atoms on a single site, assuming that the coherent dipole interaction part is zero. Each dot corresponds to an eigenstate and the color indicates the number of excitations as given in the legend. Eigenstates with the same total $F$ are degenerate. Examples of the value of $F$ for some $n_e=1$ states are given in blue font inside the plots.}
	\label{fig:n2_eigenstates}
\end{figure*}

In this section, we numerically investigate the full spectrum of eigenstates for the single-site problem. We numerically check in this way the analytical predictions of the previous section, and further provide some insights into the structure of the remaining eigenstates.

To this end, we rewrite the master equation, Eq.~(\ref{eq:HLdipoles}), as $d\hat\rho/dt = -i ( \hat{H}_\text{eff} \hat\rho - \hat\rho \hat{H}_\text{eff}^\dagger ) + \mathcal{L}_\text{rec}(\hat\rho)$ in terms of the effective non-Hermitian Hamiltonian
\begin{equation}
	\hat{H}_\text{eff} = - \sum_{i,j} \sum_{q,q'} \mathcal{G}_{q,q'}^{i,j}\, \D^+_{i,q}\, \D^-_{j,q'} ,
\label{eq:Heff}
\end{equation}
with $\mathcal{G}_{q,q'}^{i,j}=\mathcal{R}_{q,q'}^{i,j}+ i \mathcal{I}_{q,q'}^{i,j}$, and the recycling term
\begin{align}
	\mathcal{L}_\text{rec}(\hat{\rho}) =&\, 2 \sum_{i,j} \sum_{q,q'} \mathcal{I}_{q,q'}^{i,j} \D^-_{j,q'}\, \hat{\rho}\, \D^+_{i,q}.
\end{align}
The action of the effective Hamiltonian $\hat{H}_\text{eff}$ conserves the total number of excitations, $\hat n_e \equiv \sum_{i,m} \hat c^\dagger_{i,e_m}\hat c_{i,e_m}$. Therefore, its (right) eigenstates can be labelled in terms of their number of excitations $n_e$.
For a given $n_e$, the eigenvalues of the $\ket{k}_{n_e}$ eigenstate can be written as $\lambda_k = \varepsilon_k-i\gamma_k/2$, where $\varepsilon_k$ is the energy and $\gamma_k$ describes the total decay rate into the $(n_e-1)$-excited manifold.
It can also be obtained from $\gamma_k=\Tr[\mathcal{L}_\text{rec}(\ket{k}\bra{k})]$~\cite{ZollerPRA84}. Notice that this quantity only gives the decay rate into the next excited manifold, and not the total decay rate into the manifold with zero excitations.

In the following, we numerically compute the eigenstates of $\hat{H}_\text{eff}$ for $n$ atoms on a single site (we drop the index $i$ in the following).
The structure of eigenstates depends, in principle, on the geometry of the onsite trapping potential.
Recall that the onsite Green's tensor is given by $\Re G^{ii}\propto U(k_0,\boldsymbol\ell) (\mathbb{1}-3\mbf e^L_z\otimes \mbf e^L_z)$ and $\Im G^{ii}\propto \mathbb{1}$, see Eqs.~(\ref{eq:onsite_ReG}) and (\ref{eq:onsite_ImG}).
The geometry dependence enters in the prefactor $U$ and the relative orientation of the lattice and quantization axes, $\theta\equiv \angle (\mbf e_z, \mbf e^L_z)$.

For a given onsite prefactor $U$, however, the eigenvalues of $\hat H_\text{eff}$ are independent of the relative orientation of lattice and quantization axis ($\theta$), and the eigenstates for different orientations are related by rotations with respect to $\theta$. This is because one can always define the atomic basis with the quantization axis pointing along the lattice direction. The obtained eigenstates can then be written in terms of a basis with a different quantization axis by applying a rotation. However, the eigenvalues remain untouched in this operation. The situation is of course different if an explicit symmetry-breaking magnetic field is applied, which then singles out a particular direction.

Except for the dark states and other particular cases, the energies and decay rates generally depend on the value of the prefactor $U$. For simplicity, we will consider in the following the limit of an isotropic local lattice potential such that the coherent part of the onsite dipolar interactions is zero, $U\equiv0$. In this case, the effective Hamiltonian is given by
\begin{align}
	\hat{H}_\text{eff} \propto&\, \sum_{q} \D^+_q\,\D^-_q.
\label{eq:Heff_C0}
\end{align}
Note that the only terms contributing to the dynamics are those with $q=q'$ and hence it conserves the total magnetic number,\footnote{The total magnetic number $M$ is also conserved for $U\neq0$ when $\mbf e^L_z\parallel \mbf e_z$. In that case, the terms $\mbf e^{*T}_q (\mbf e^L_z\otimes \mbf e^L_z) \mbf e_{q'}$ originating from the coherent part are nonzero only for $q=q'=0$, which conserves the total $M$.}
such that eigenstates can be labelled by both $n_e$ and $M$.

The expression $\sum_{q} \D^+_q\,\D^-_q$ appearing in Eq.~(\ref{eq:Heff_C0}) is a scalar operator under rotations, because it can be written as the scalar product of two spherical operators (see App.~\ref{app:dipole_op}). This implies that this operator commutes with the total angular momentum operator $\hat{\mbf F}$, and hence all its eigenstates can be labelled by $F$. Therefore, the eigenstates of the Hamiltonian in Eq.~(\ref{eq:Heff_C0}) are given by the total angular momentum eigenstates. Moreover, given that the Hamiltonian does not depend on the quantization axis, we expect a degeneracy of all $M$ states corresponding to the same angular momentum quantum numbers.

\begin{figure*}[t!]
\centering
	\includegraphics[width=.75\textwidth]{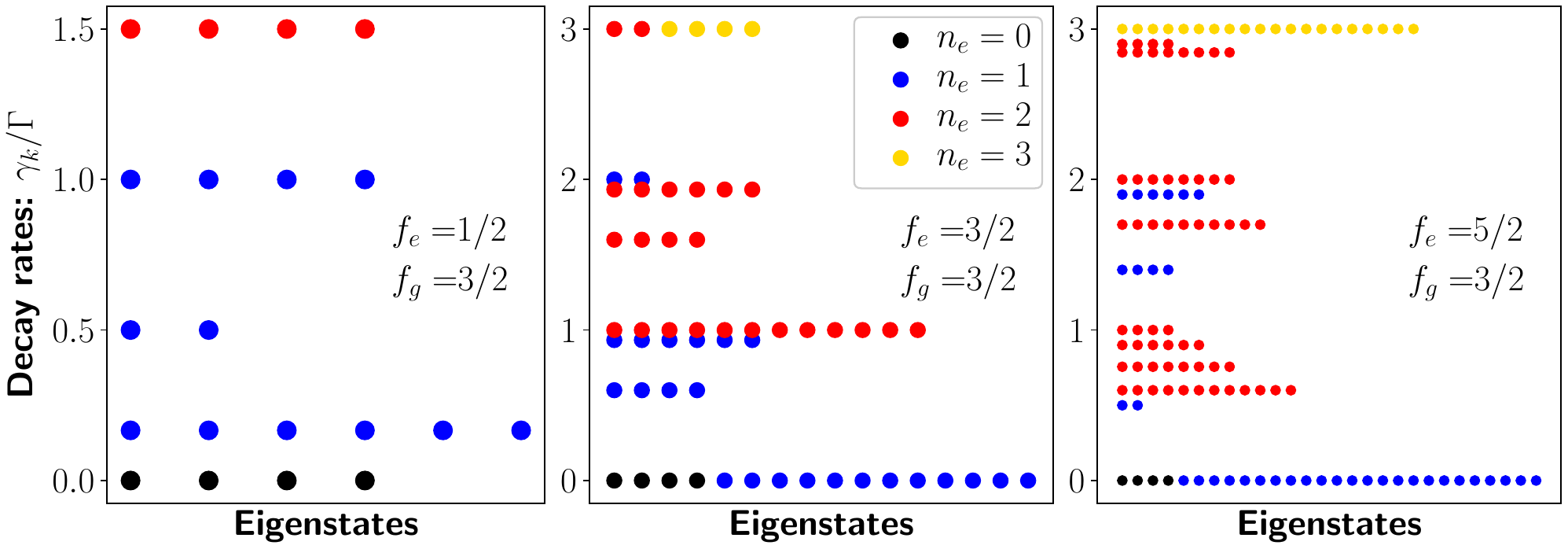}
	\caption{\textbf{Eigenstates for $n=3$.} Decay rates $\gamma_k/\Gamma$ of all eigenstates $\ket{k}$ for different level structures (indicated inside the plots) for $n=3$ atoms on a single site, assuming that the coherent dipole interaction part is zero. Each dot corresponds to an eigenstate and the color indicates the number of excitations as given in the legend.}
	\label{fig:n3_eigenstates}
\end{figure*}

\begin{figure*}[t!]
\centering
	\includegraphics[width=.75\textwidth]{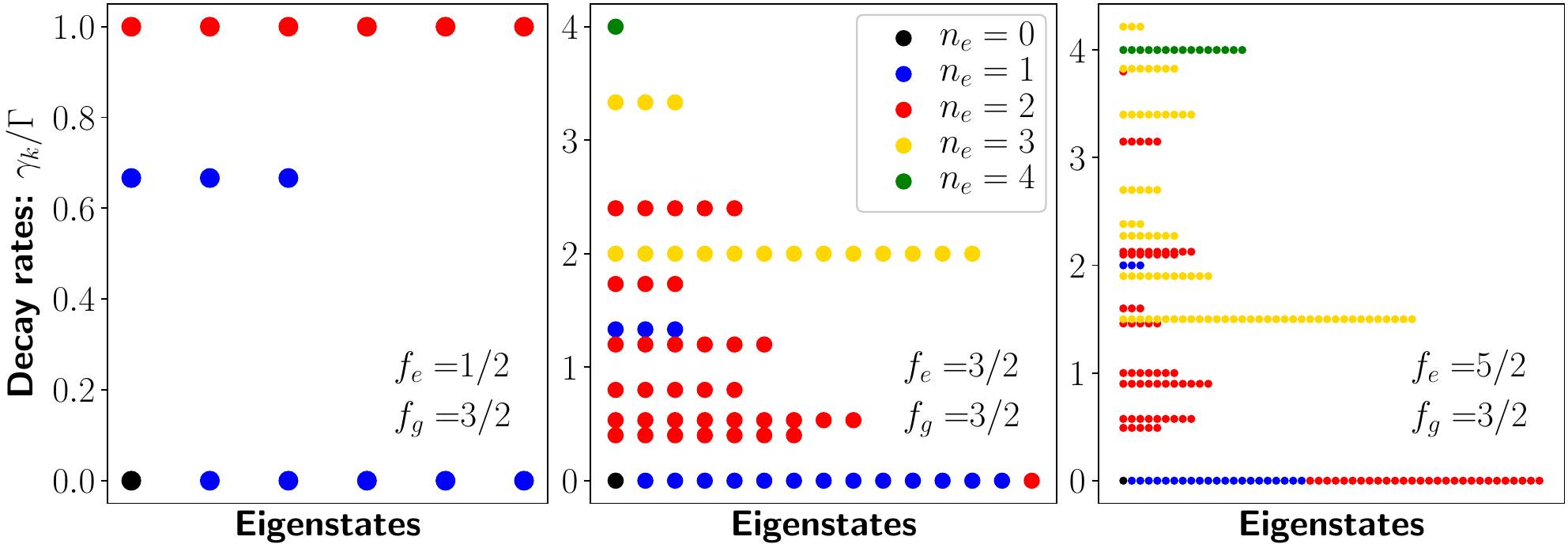}
	\caption{\textbf{Eigenstates for $n=4$.} Decay rates $\gamma_k/\Gamma$ of all eigenstates $\ket{k}$ for different level structures (indicated inside the plots) for $n=4$ atoms on a single site, assuming that the coherent dipole interaction part is zero. Each dot corresponds to an eigenstate and the color indicates the number of excitations as given in the legend.}
	\label{fig:n4_eigenstates}
\end{figure*}

Figure~\ref{fig:n2_eigenstates} shows for filling $n=2$ the decay rates $\gamma_k$ of the different eigenstates $k$ for the multi-$\Lambda$ (left), multi-$\square$ (top) and multi-$V$ (bottom) level structures. We provide examples for $f_g=1/2$ relevant for $^{171}$Yb, $f_g=5/2$ relevant for $^{173}$Yb, and $f_g=9/2$ relevant for $^{87}$Sr. The different colors correspond to different numbers of excitations $n_e$.

In all cases shown in Fig.~\ref{fig:n2_eigenstates} the number of dark states ($\gamma_k$) matches the analytical prediction of the previous section, Sec.~\ref{ssec:darks_n2}, i.e.~no dark states for the multi-$\Lambda$, 1 dark state for the multi-$\square$, and $2f+1$ dark states for the multi-$V$ level structure. The remaining eigenstates form various degenerate manifolds with the same decay rate, either sub- or superradiant. The multiplicity matches perfectly well the Zeeman multiplicity of the $\ket{F,M}_{ge}$ states. This allows to straightforwardly associate the different eigenstates and decay rates with the $\ket{F,M}_{ge}$ states. Some examples of the total $F$ of the eigenstates are given (in blue font) inside the plots.
Apart from this, the $n_e=0$ and $n_e=2$ eigenstates trivially decay with $0$ and $2\Gamma$, respectively.

Note that due to Pauli blocking some states with $\gamma_k<\Gamma$ may actually be superradiant, in the sense that interference from emission is constructive. For example, for $f_e=f_g=1/2$ the eigenstate $\ket{1,0}_{ge}$ decays with $\gamma_k=2/3\Gamma$. This state is, however, superradiant compared to the decay $\tilde{\Gamma}=1/3\Gamma$ obtained for the states $\ket{g_{\mp1/2}e_{\pm1/2}}$ when dipole interactions are ignored and only Pauli exclusion is taken into account . The prefactor results from a Clebsch-Gordan coefficient.

Figure~\ref{fig:n3_eigenstates} shows for filling $n=3$ the decay rates of the eigenstates for the multi-$\Lambda$, $\square$, and $V$ level structures with $f_g=3/2$. Again, the number of dark states predicted in the analytical section, Sec.~\ref{ssec:darks_n3}, matches the number of dark states found numerically in Fig.~\ref{fig:n3_eigenstates}, i.e.~none for multi-$\Lambda$, 12 for multi-$\square$, and 24 for multi-$V$. We also checked that the dark states predicted for $f_g=5/2$ match the numerical results. All the remaining eigenstates are again grouped into degenerate sets of states and their degeneracy is consistent with the $2F+1$ degeneracy expected for states of same total angular momentum $F$.

The case of states with the same $F$ but different $f_{12}$ is interesting. For multi-$\square$ ($f_g=3/2$) we have two sets of single-excitation states with $F=3/2$ for $f_{12}=0$ and $2$, respectively, see Eq.~(\ref{eq:n3square_states_gge}). These states separate into two sets of 4 eigenstates each, made up of superpositions of $f_{12}=0,2$ states. One of these sets is dark, as argued in Sec.~\ref{ssec:darks_n3}. The other 4 superposition states are nondark and degenerate, as seen from the 4 blue dots at around $2/3\Gamma$ in Fig.~\ref{fig:n3_eigenstates}. The same happens for multi-$V$, where the two sets of single-excitation $F=5/2$ states split into a dark subset and a degenerate subset at around $2\Gamma$, see Fig.~\ref{fig:n3_eigenstates}.

Figure~\ref{fig:n4_eigenstates} shows for filling $n=4$ the decay rates of the eigenstates for the multi-$\Lambda$, $\square$, and $V$ level structures with $f_g=3/2$. The number of dark states again agrees with the analytical discussion of Sec.~\ref{ssec:darks_n3}.
In particular, notice that in this case the multi-$\Lambda$ system has for the first time dark states.

%%%%%%%%%%%%%					%%%%%%%%%%%%%
%%%%%%%%%%%%%		SECTION		%%%%%%%%%%%%%
%%%%%%%%%%%%%					%%%%%%%%%%%%%

\section{Preparation\label{sec:preparation}}

We discuss in this section different possibilities to coherently excite the single-site dark states presented in Sec.~\ref{sec:darks} starting with the atoms in the ground state. In particular, we extend the discussion on the Raman excitation scheme presented in Ref.~\cite{PineiroArxiv1907}, and consider another method based on lifting the Zeeman sublevels degeneracy using magnetic fields or off-resonant dressing beams. In the following, we consider only doubly-occupied sites, $n=2$, since they will be experimentally most relevant, and consider only a single-site.

Before starting it is instructive to note that, starting with both atoms in the ground manifold, the atoms can not be directly excited to a dark state using simply a laser. This is because the matrix element of the laser Hamiltonian connecting dark and ground states is zero, for the same reason that makes dark states dark. More specifically, the laser Hamiltonian is given by $\hat{H}_L=-\sum_q\big( \Omega_q\,\D^+_q+\text{h.c.} \big)$ with Rabi frequency $\Omega_{q} \equiv \Omega \left( \mbf e^*_{q} \cdot \boldsymbol\epsilon_L \right)$ and laser polarization $\boldsymbol\epsilon_L$. Since the dark states $\ket{D}$ considered here fulfill the condition (\ref{eq:dark_condition}), this implies that $\bra{D}\hat{H}_L\ket{G}=0$ for any ground state $\ket{G}$.

%%% ----------------------------------------------------------------------------
%%%				Subsection
%%% ----------------------------------------------------------------------------

\subsection{Raman excitation}

\begin{figure*}[t!]
\centering
	\includegraphics[width=\textwidth]{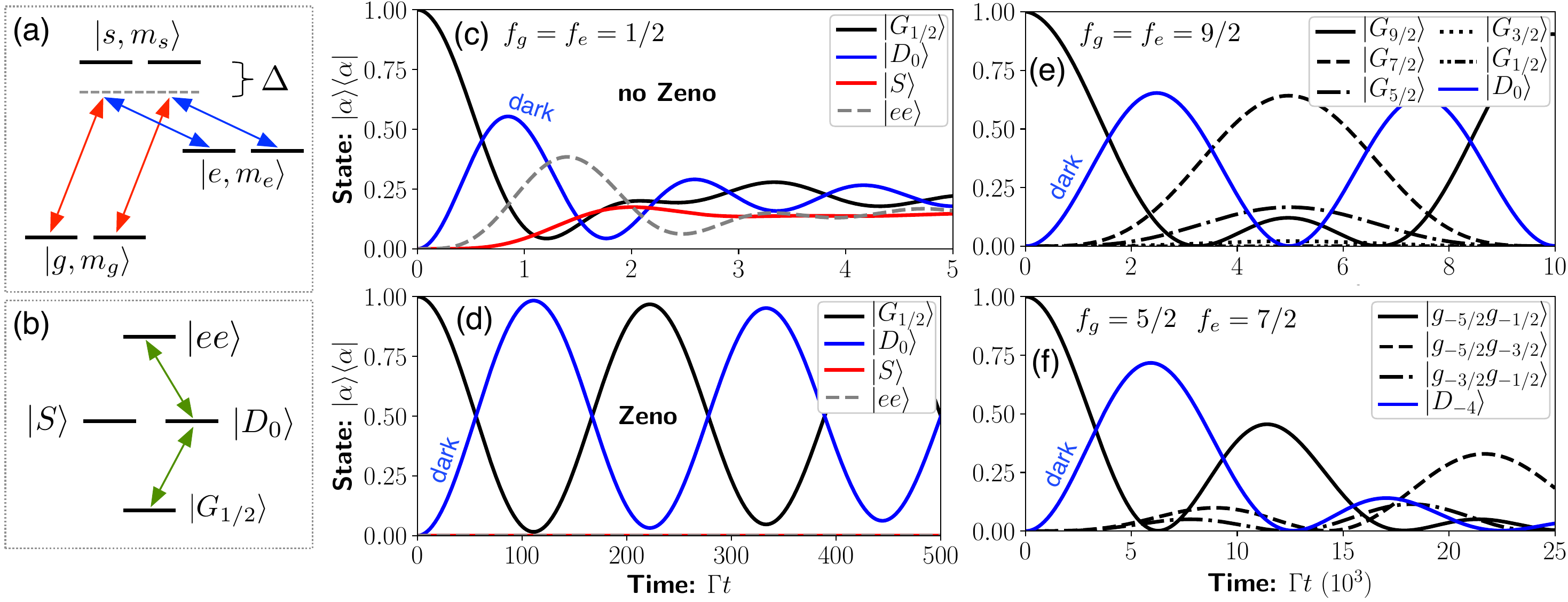}
	\caption{\textbf{Raman excitation scheme.}
	\emph{Left:} Sketch of Raman excitation for $f_g=f_e=f_s=1/2$ and $\mbf e_z$ polarized light in (a) single-particle and (b) eigenstate basis.
	\emph{Middle:} Excitation of dark state $\ket{D_0}_{\{1/2,1/2\}}$ for $f_g=f_e=f_s=1/2$ level structure with (c) $\Omega^\text{eff}=3\Gamma$ (no Zeno effect) and (d) $\Omega^\text{eff}=0.03\Gamma$, which leads to Zeno suppression of nondark states.
	\emph{Right:} (e) Excitation of dark state $\ket{D_0}_{\{9/2,9/2\}}$ for $f_g=f_e=f_s=9/2$ structure with $\Omega^\text{eff}=0.001\Gamma$. (f) Excitation of dark state $\ket{D_{-4}}_{\{5/2,7/2\}}$ for $f_g=f_s=5/2$, $f_e=7/2$ level structure with $\Omega^\text{eff}=0.001\Gamma$.}
	\label{fig:Raman}
\end{figure*}

The Raman excitation scheme consists in coupling both $g$ and $e$ states off-resonantly to an intermediate state $s$, for which we assume a total angular momentum $f_s$, $m_s\in[-f_s,f_s]$, frequency $\omega_s$, and decay rate $\Gamma_s$, see Fig.~\ref{fig:Raman}(a). The Hamiltonian is given by (in units of $\hbar$)
\begin{align}
	\hat{H}^{(1,2)}_L =&\, -\sum_{m,n} \left[ \Omega^{(1)}_{s_mg_n}\,\hat{\sigma}_{s_mg_n} + \Omega^{(2)}_{s_me_n}\,\hat{\sigma}_{s_me_n} + \text{h.c.} \right] \nonumber\\
	&\, -\Delta \sum_m \hat\sigma_{s_ms_m},
\end{align}
where $\Omega^{(l)}_{a_mb_n}=\Omega^{(l)}e^{i\phi_l}\langle f_b\, n; 1\, m-n | f_a\, m \rangle (\mbf e^*_{m-n}\cdot\boldsymbol\epsilon^{(l)}_L)$, $\boldsymbol\epsilon^{(l)}_L$ is the polarization of the $l$-laser, $\Omega^{(l)}$ is the Rabi frequency, $\phi_l=\mbf k^{(l)}_L\cdot\mbf r_i$ is the phase of the laser at array site $i$, $\omega_L^{(l)}=c k^{(l)}_L$ is the laser frequency, and the first sum runs only over values with $|m-n|\leq1$. The Hamiltonian is given in a frame where the states $e$ rotate with $\omega_0$ and the states $s$ with $\omega_L^{(1)}$ (the frequency of the $g$-$s$ laser). In this frame, the $s$ states are detuned by $\Delta = \omega^{(1)}_L-\omega_s$. Notice that a site dependence enters in the phase $\phi_l$, but we suppress the $i$ index for notational simplicity.

When the detuning is large, $|\Delta|\gg\Omega^{(1)},\Omega^{(2)},\Gamma_s$, the intermediate state can be adiabatically eliminated~\cite{JerkeCanada2007} and the resulting effective Hamiltonian reads
\begin{align}
	\hat{H}_\text{Raman} = \sum_{m,n} \left[ \Omega^{\text{eff}}_{mn}\, \hat\sigma_{e_mg_n} + \text{h.c.} \right]
\end{align}
with effective Rabi coupling $\Omega^{\text{eff}}_{mn} \equiv \sum_k \Omega^{(2)*}_{s_ke_m} \Omega^{(1)}_{s_kg_n} /\Delta$.
The reason why this Raman Hamiltonian will allow us to couple to the dark state is that the effective Rabi couplings $\Omega^\text{eff}_{mn}$ have, in general, a different structure than the single-laser couplings.

As an example, consider $f_g=f_e=f_s=1/2$ and let us start with both atoms in the ground state, $\ket{G_{1/2}}\equiv\ket{g_{-1/2}g_{1/2}}$.
Using a single $\mbf e_z$-polarized laser, the corresponding Hamiltonian is given by $\hat{H}_L \propto \sum_{m} C^{0}_m \left[ \hat\sigma_{e_mg_m} + \text{h.c.} \right]$, which couples the initial state to the superradiant state $\ket{S}\equiv \left( \ket{g_{1/2}\,e_{-1/2}} + \ket{g_{-1/2}\,e_{1/2}} \right)/\sqrt{2}$.
This is because for this level structure we have $C^{0}_{1/2}=-C^{0}_{-1/2}$.
In contrast, if we use the Raman scheme with two $\mbf e_z$-polarized lasers, we get $\hat{H}_L^{\text{eff}} \propto \sum_{m} (C^{q=0}_m)^2 \left[ \hat\sigma_{e_mg_m} + \text{h.c.} \right]$, which instead couples the ground state to (only) the dark state $\hat{H}^\text{eff}_L\ket{G_{1/2}}\propto\ket{D_0}_{\{\frac{1}{2},\frac{1}{2}\}}$. Note that the Raman scheme essentially changes the symmetry of the state as compared to the single laser, due to the square $(C^{0}_{1/2})^2=(C^{0}_{-1/2})^2$.

Figure~\ref{fig:Raman}(c) shows an example of this for a relatively large effective Rabi coupling $\Omega^\text{eff}=3\Gamma$, where we defined $\Omega^\text{eff}\equiv\Omega^{(1)}\Omega^{(2)}/\Delta$ and set $\phi_l=0$. As it can be seen, the lasers manage to excite the dark state (blue line) up to an occupation $\sim0.5$. However, the same lasers also couple the dark state to the double-excited state $\ket{ee}\equiv\ket{e_{-1/2}e_{1/2}}$ (gray line), which subsequently decays, e.g.~into the superradiant state $\ket{S}$ (red line), see Fig.~\ref{fig:Raman}(b). The population of the $\ket{ee}$ state can be suppressed by choosing a small effective Rabi coupling, $\Omega^\text{eff}\ll\Gamma$. Due to the quantum Zeno effect, the doubly-excited state is then only populated at a rate $\sim(\Omega^\text{eff})^2/\Gamma$. Figure~\ref{fig:Raman}(d) shows that high-contrast coherent oscillations between ground and dark state are obtained for $\Omega^\text{eff}=0.03\Gamma$.

The same method can be used to address other dark states in other level structures. For this, one needs to choose (i) an initial ground-state $\ket{g_mg_n}$ and (ii) an intermediate state and laser polarizations such that the coupling to the dark state is nonzero. Figure~\ref{fig:Raman}(e) shows an example for $f_g=f_s=f_e=9/2$, starting with $\ket{G_{9/2}}=\ket{g_{-9/2}g_{9/2}}$ and using $\mbf e_z$ polarized lasers. Using again a small Rabi coupling to suppress population in other non-dark states, we obtain coherent oscillations of the dark state $\ket{D_0}_{\{9/2,9/2\}}$. In this case, however, other ground states $\ket{G_{M<9/2}}\equiv\ket{g_{-M}g_{M}}$ get excited along the way, such that we do not achieve a full inversion into the dark state. Figure~\ref{fig:Raman}(f) shows another example for $f_g=f_s=5/2$, $f_e=7/2$, starting with $\ket{g_{-5/2}g_{-1/2}}$ and using $\boldsymbol\epsilon_L^{(1)}=\mbf e_z$, $\boldsymbol\epsilon_L^{(2)}=\mbf e_+$. Again, we can achieve a large (but not unit) occupation of the dark state $\ket{D_{-4}}_{\{5/2,7/2\}}$. As opposed to before though, the process does not rephase.

We note that, in reality, energy shifts from collisions will make some transitions off-resonant, so the Zeno effect may not be always necessary. This is in particular true of double-excited states.
Apart from this, we emphasize that the aim of this section was to coherently prepare a given dark state. If the conditions for Zeno suppression are not experimentally feasible, the dark state can also be prepared in a probabilistic manner using a larger Rabi coupling.

%%% ----------------------------------------------------------------------------
%%%				Subsection
%%% ----------------------------------------------------------------------------

\subsection{Zeeman excitation\label{ssec:zeeman_excit}}

\begin{figure*}[t!]
\centering
	\includegraphics[width=\textwidth]{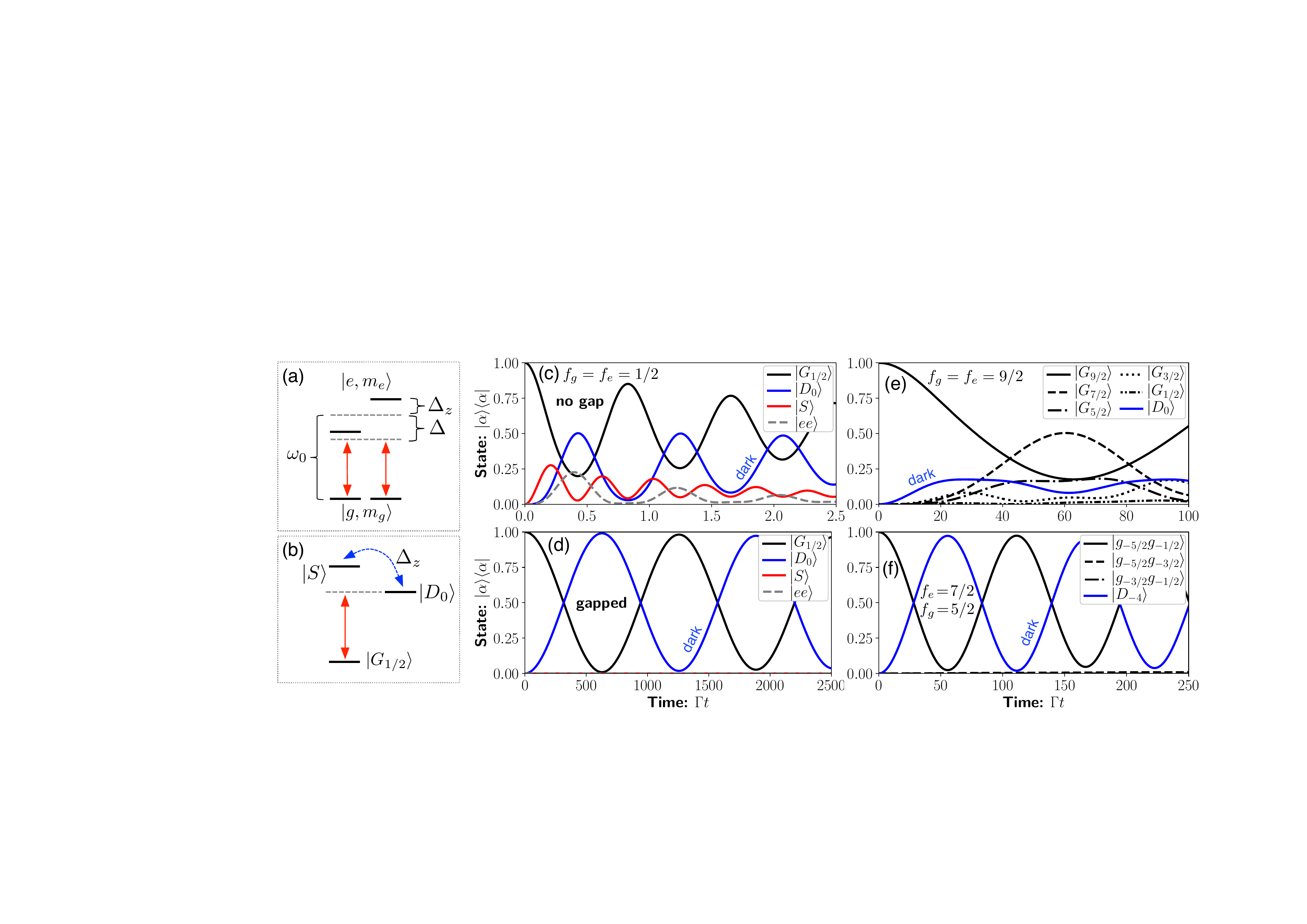}
	\caption{\textbf{Zeeman excitation scheme.} \emph{Left:} Sketch of Zeeman excitation for $f_g=f_e=1/2$ in (a) single-particle and (b) eigenstate basis. Superradiant state $\ket{S}$ and dark state $\ket{D_0}$ are defined in text. \emph{Middle:} Excitation of $\ket{D_0}_{\{1/2,1/2\}}$ for $f_g=f_e=1/2$ level structure starting from $\ket{G_{1/2}}$ with (c) $\Delta_z=10\Gamma$, $\Omega=5\Gamma$, $\Delta=0$, $U=0$ (no gap), and with (d) $\Delta_z=1\Gamma$, $\Omega=\sqrt{3/8}\Gamma$, $\Delta=0$, $U=100\Gamma$ (gapped). \emph{Right:} Excitation of different dark states: (e) $f_g=f_e=9/2$ starting from $\ket{G_{9/2}}$ with $\Delta_z=1\Gamma$, $\Omega=2.5\Gamma$, $\Delta=0$, $U=100\Gamma$ (f) $f_g=5/2$, $f_e=7/2$ starting from $\ket{g_{-5/2}g_{-1/2}}$ with $\Delta_z=3\Gamma$, $\Omega=1.5\Gamma$, $\Delta=-6.99\Gamma$, $U=100\Gamma$.}
	\label{fig:Zeeman}
\end{figure*}

Another alternative to coherently excite a dark state is to add a detuning between the Zeeman sublevels of the $g$ and/or $e$ manifolds, see Fig.~\ref{fig:Zeeman}(a). This can be achieved by off-resonant dressing, or applying external magnetic fields. When the Zeeman sublevels are not degenerate, the dark states are not perfect eigenstates anymore, but instead couple to other bright states with a strength proportional to the Zeeman splitting, see Fig.~\ref{fig:Zeeman}(b). Shining on a laser with appropriately chosen characteristics will then allow to couple to the dark state through a sort of Raman transition with the bright state as intermediate state.
This is particularly well-suited when there is a large energy gap between dark and bright states, due to e.g.~collisional shifts, because then the bright state is only virtually occupied.

We consider the atoms to be addressed by a laser with detuning $\Delta$ from resonance, and subject to an external magnetic field. We assume that the $g$-factor of the ground state is negligible and that the Zeeman splitting between subsequent $e$-levels is $\Delta_z$. Thus, the total Hamiltonian in the laser's rotating frame is (in units of $\hbar$)
\begin{align}
	\hat{H}_\text{Zeeman}=&\, -\sum_{m,n} \left[ \Omega_{e_mg_n}\,\hat{\sigma}_{e_mg_n} + \text{h.c.} \right] + \sum_m \Delta_m\, \hat\sigma_{e_me_m},
\label{eq:Zeeman_H}
\end{align}
with $\Omega_{e_mg_n}=\Omega\, C^{m-n}_{n} (\mbf e^*_{m-n}\cdot\boldsymbol\epsilon_L)$ and $\Delta_m=m\Delta_z-\Delta$.
Furthermore, we emulate the possibility of having energy shifts between dark and bright states by varying the strength $U$ of the coherent dipolar interactions [see Eq.~(\ref{eq:onsite_ReG}) and App.~\ref{app:onsite}].

As a specific example, we consider again the case $f_g=f_e=1/2$ with $\mbf e_z$ polarized light. The laser couples the ground state $\ket{G_{1/2}}$ to the superradiant bright state $\ket{S}$ with a strength $\tilde\Omega\equiv-\Omega\sqrt{2/3}$ due to Clebsch-Gordan coefficients, whereas the Zeeman detuning couples $\ket{S}$ with $\ket{D_0}$ with a strength $\Omega_z=\Delta_z/2$. Thus, the Hamiltonian (\ref{eq:Zeeman_H}) in the $\{\ket{G_{1/2}},\ket{S},\ket{D_0}\}$ basis reads
\begin{align}
	\hat{H}_\text{Zeeman} + \hat{H}_\text{shifts} =&\, \begin{pmatrix} 0 & \tilde\Omega & 0 \\ \tilde\Omega & \epsilon_S - \Delta & \Omega_z \\ 0 & \Omega_z & \epsilon_D-\Delta \end{pmatrix}.
\end{align}
Here, we included the energy shifts $\epsilon_S$ and $\epsilon_D$ of the superradiant and dark states. Note that if these shifts came from the coherent dipole interaction alone, then we would have $\epsilon_D=0$.

Figure~\ref{fig:Zeeman}(c) shows the result of this excitation scheme assuming no energy shifts ($U=0$), $\epsilon_S=\epsilon_D=0$. Here, we chose $\Delta=0$ to make the dark state resonant, and a large Zeeman splitting $\Delta_z=10\Gamma$ with $\Omega=5\Gamma$. Although the dark state gets excited up to about $0.5$, the superradiant state and the double-excited state also do. Their population can be suppressed if there is an energy gap, $\delta_\text{gap}\equiv \epsilon_S-\epsilon_D$, between the dark and the superradiant state. Figure~\ref{fig:Zeeman}(d) shows an example of this for $U=100\Gamma$, $\Delta_z=1\Gamma$, $\Omega=\sqrt{3/8}\Gamma$. The latter value is chosen such that $|\tilde\Omega|=|\Omega_z|$.

Figures~\ref{fig:Zeeman}(e) and (f) show further examples of this excitation scheme applied to more complicated cases: (e) $\ket{D_0}$ for $f_g=f_e=9/2$ and (f) $\ket{D_{-4}}$ for $f_g=5/2$, $f_e=7/2$. In these cases, the initial ground state couples to more than one bright eigenstates, which in turn couple to the dark state. Because of this, estimating the optimal values for the parameters becomes more complicated. Nevertheless, Figs.~\ref{fig:Zeeman}(e) and (f) show that with appropriately chosen parameters, we can reach a large occupation of the dark state. In Fig.~\ref{fig:Zeeman}(e) the occupation does not reach 1, because other ground-states get excited along the way, similarly to the Raman scheme in the previous section [see Fig.~\ref{fig:Raman}(e)]. In Fig.~\ref{fig:Zeeman}(f) we note that a detuning of $\Delta=-2.33\Delta_z$ was used to compensate the energy shift of the dark state due to the Zeeman term.

%%%%%%%%%%%%%					%%%%%%%%%%%%%
%%%%%%%%%%%%%		SECTION		%%%%%%%%%%%%%
%%%%%%%%%%%%%					%%%%%%%%%%%%%

\section{Implementation\label{sec:implementation}}

Alkaline-earth(-like) atoms are promising candidates to realize the multilevel dark states presented here due to their lack of hyperfine coupling in the ground state, which has total electronic angular momentum $J=0$. Examples of possible isotopes currently used in experiments~\cite{GobanNature2018,FollingNature2014} include $^{171}$Yb, $^{173}$Yb, or $^{87}$Sr, which have nuclear spins $I=1/2$, $I=5/2$ and $I=9/2$, respectively. For this, one possibility would be to use the clock transition $g= {}^1\text{S}_0$ to $e= {}^3\text{P}_0$, which has a linewidth $\Gamma\sim \text{mHz}$. As intermediate states for the Raman state preparation one could use $s= {}^1\text{P}_1$, $s= {}^1\text{S}_1$, $s= {}^1\text{D}_2$, or $s= {}^3\text{D}_2$.

However, because of the generally ultra-long lifetime of the bare ${}^3\text{P}_0$ state, observing an enhancement of this lifetime due to subradiance will be challenging, especially because of other competing decay mechanisms such as light scattering from the laser beams used to trap the atoms~\cite{HutsonYePRL123}. This limits the lifetime of any state, including the dark state, due to dressing with higher-lying short-lived electronic levels.
To circumvent this issue, we propose instead to prepare the dark state in the ${}^1\text{S}_0$-${}^3\text{P}_0$ manifold and then off-resonantly dress the excited state with a faster decaying transition such as ${}^1\text{P}_1$ or ${}^3\text{P}_1$, similarly to the proposal in Ref.~\cite{SantraGreenePRL94}. This would not only increase the effective decay rate of the dark state, but also allow to controllably switch on and off the decay, which would leave ample time to do the measurement.
Notice that the interference mechanism needed for subradiance would be preserved by such an off-resonant dressing, as long as the intermediate state predominantly decays to ${}^1\text{S}_0$, where the other atom is.

Alternatively, one can aim at preparing the dark state in a faster decaying transition such as ${}^3\text{P}_1$ with $\Gamma\sim1-100\,\text{kHz}$. In this case, the dark state could also be prepared using the ${}^3\text{P}_0$ state as intermediary for the Raman scheme. The advantage of this would be that due to the long lifetime of the ${}^3\text{P}_0$ state, the preparation procedure could be accomplished in two steps by applying two subsequent $\pi$-pulses. Another possibility would be to prepare the dark state in the ${}^1\text{P}_1$ state, except that $\Gamma\sim\text{MHz}$ strong dipolar interactions may induce fast mixing with higher motional bands and the atoms in the excited state will not be trapped, as noted in Ref.~\cite{SantraGreenePRL94}.

%%%%%%%%%%%%%					%%%%%%%%%%%%%
%%%%%%%%%%%%%		SECTION		%%%%%%%%%%%%%
%%%%%%%%%%%%%					%%%%%%%%%%%%%

\section{Experimental limitations\label{sec:limitations}}

One important limitation to the lifetime of the dark state will be set by the effective decay rate to higher motional bands. In the Lamb-Dicke regime assumed here ($\eta\ll1$), the decay to the next excited band is given by $\Gamma_\text{eff}\sim\eta^2\Gamma$~\cite{ZollerSandnerPRA2011}. This effect can in principle be suppressed by increasing the intensity of the trapping laser. Note, however, that this will also lead to larger light scattering effects.

Another source of decay are non-vanishing stray magnetic fields. As explained above, such terms generally couple dark and bright states to each other. While this can be used for preparation (see Sec.~\ref{ssec:zeeman_excit}), once prepared a non-vanishing Zeeman detuning $\Delta_z$ will induce an effective decay rate on the dark state. Fortunately, if this detuning is much smaller than the linewidth, $\Delta_z\ll\Gamma$, the quantum Zeno effect will suppress this effect as $\Gamma_\text{eff}\sim\Delta_z^2/\Gamma$. Alternatively, the Zeeman coupling can be suppressed if the dark state is energetically well-separated from the bright state. This is, for instance, the case in the ${}^3\text{P}_0$ state, where energy shifts due to collisions are $\sim\text{kHz}$~\cite{GobanNature2018,FollingNature2014}.

Collisions can also induce mixing between dark and bright states. However, if the atoms are tightly trapped such that they are confined to be in the motional ground state, the complexity of the collisions is greatly reduced.
In particular, since the dark states are eigenstates of the total angular momentum, they should be good eigenstates of collisions too, as long as collisions are dominated by the long-range part of the potential. The effect of the short-range interaction part, which is beyond the scope of this work, is generally hard to estimate~\cite{ReichenbachPRL99,ReichenbachPRA80,ReichenbachPRA82} and ultimately should be tested experimentally.

%%%%%%%%%%%%%					%%%%%%%%%%%%%
%%%%%%%%%%%%%		SECTION		%%%%%%%%%%%%%
%%%%%%%%%%%%%					%%%%%%%%%%%%%

\section{Conclusions\label{sec:conclusions}}

In this work, we have studied the subradiance properties of $n\geq2$ multilevel fermions loaded into a single trap (e.g.~an optical lattice site or a tweezer), thus extending our previous work~\cite{PineiroArxiv1907}. The atoms are assumed to occupy the lowest motional band and interact with each other dominantly via dipolar (coherent and incoherent) exchange interactions. The multilevel nature combined with fermion statistics and quantum interference gives rise in this system to a set of dark (perfectly subradiant) eigenstates.
While the focus of the work was on single sites, multi-site dark states on optical lattices or tweezers can be prepared by creating product states where atoms in each site are either in the ground or in an excited single-site dark state (or in a superposition of them).

We have found and characterized all single-site dark states appearing for different fillings $n$ and internal level structures $f_g\leftrightarrow f_e$. We have shown that these dark states are related to eigenstates of the total angular momentum of the $n$ atoms, and we have given generic prescriptions to identify dark states by counting their respective decay channels, using dipole selection rules, and the properties of dipole matrix elements.
Moreover, we have seen that the full eigenstate structure of the single-site problem corresponds to total angular momentum states when the coherent dipolar interaction part is zero. This shows the usefulness of the total angular momentum basis to describe multilevel systems.

We proposed two different schemes to coherently prepare the dark states with close to $100\%$ fidelity in some cases. The first approach is based on a Raman like transition, which allows to couple ground and dark states using the properties of Clebsch-Gordan coefficients. The second approach instead makes use of magnetic-field-induced energy shifts between Zeeman sublevels to couple bright and dark states.

These multilevel dark states can be implemented using alkaline-earth atoms in optical lattices or tweezers and can find applications in different quantum technologies. Superpositions of ground and dark states can be used e.g.~as logical clock states with vanishingly small decay rate for quantum metrology. In particular, this opens the door to building atomic optical clocks on internal level transitions that are not naturally long lived. A big advantage of using these multilevel dark states for clocks would be that such states would not suffer from dipolar interaction shifts (since they are zero eigenstates thereof), which will potentially limit the accuracy of current 3D lattice clocks.
Apart from this, ground and dark states could also be used as the basis for decoherence-free qubits in quantum information, for quantum simulation, and to build interesting quantum optical devices.

\begin{acknowledgments}
We thank D.~Barberena, I.~Kimchi, S.~F\"olling, P.~Julienne, J.~P.~D'Incao, C.~Sanner, J.~Ye, J.~K.~Thompson, A.~Kaufman, M.~Perlin for useful discussions on the topic.
This work is supported by the AFOSR grant FA9550-18-1-0319 and its MURI Initiative, by the DARPA and ARO grant W911NF-16-1-0576,  the ARO single investigator award W911NF-19-1-0210,  the NSF PHY1820885, NSF JILA-PFC PHY-1734006 grants, and by NIST.
\end{acknowledgments}

\vfill

\appendix

%%%%%%%%%%%%%					%%%%%%%%%%%%%
%%%%%%%%%%%%%		SECTION		%%%%%%%%%%%%%
%%%%%%%%%%%%%					%%%%%%%%%%%%%

\section{Derivation of master equation\label{app:master_eq_deriv}}

%%% ----------------------------------------------------------------------------
%%%				Subsection
%%% ----------------------------------------------------------------------------

\subsection{Light-matter Hamiltonian}

We derive in this section the multilevel dipolar master equation, Eq.~(\ref{eq:HLdipoles}), starting from the light-matter Hamiltonian of Eqs.~(\ref{eq:Hatom}), (\ref{eq:Hfield}), and (\ref{eq:Hatomfield}), following Refs.~\cite{GrossHarochePRep1982,JamesPRA47,LehmbergPRA2}, see also Ref.~\cite{ChangYeLukinPRA69}.
We recall that the light-matter Hamiltonian is given by
\begin{equation}
	\hat{H}_\text{tot} = \hat{H}_\text{atom} + \hat{H}_{\text{field}} + \hat{H}_\text{af},
\end{equation}
with
\begin{align}
	\hat{H}_\text{atom} =&\, \int d\mbf r\, \sum_{m} \hbar\omega_{e_m} \hat\sigma_{e_me_m}(\mbf r) + \sum_{n} \hbar\omega_{g_n} \hat\sigma_{g_ng_n}(\mbf r),\\
	\hat{H}_{\text{field}} =&\, \sum_{\mbf k,\lambda} \hbar\omega_k \left( \hat{a}^\dagger_{\mbf k,\lambda} \hat{a}_{\mbf k,\lambda} + \frac{1}{2} \right),\\[5pt]
	\hat{H}_\text{af} =&\, - \sum_{m,n} \sum_{\mbf k,\lambda} \int d\mbf r\, \mathrm{g}_k \Big( \mbf d_{e_m g_n}\, \hat\sigma_{e_mg_n}(\mbf r) + \text{h.c.} \Big) \nonumber\\
	&\,\qquad\qquad\qquad\qquad \cdot\Big( \boldsymbol\epsilon_{\mbf k,\lambda} \hat{a}_{\mbf k,\lambda} e^{i\mbf k\mbf r} + \text{h.c.} \Big).
\end{align}
Here, we consider arbitrary energies $\omega_{e_m}>\omega_{g_n}$ to include the possibility of having nondegenerate Zeeman sublevels.

We start by switching to the interaction picture with $\hat{H}_0\equiv \hat{H}_{\text{atom}}+\hat{H}_{\text{field}}$, where the equation of motion for the full atom-field density matrix $\hat{\rho}_{\text{af}}$ is given by
\begin{equation}
	\frac{d\hat{\rho}_{\text{af}}}{dt} = \frac{1}{i\hbar} \left[ \hat{V}(t) , \hat{\rho}_{\text{af}} \right],
\label{eq:intpic_schroed}
\end{equation}
with the interaction part $\hat V(t) = e^{i\hat{H}_0 t/\hbar} \hat{H}_{\text{af}}\, e^{-i\hat{H}_0 t/\hbar}$. It is given by
\begin{align}
	\hat{V}(t) =&\, - \sum_{m,n} \sum_{\mbf k,\lambda} \int d\mbf r\, \mathrm{g}_k \Big( \mbf d_{e_m g_n}\, \hat\sigma_{e_mg_n}(\mbf r)\,e^{it\omega_{mn}} + \text{h.c.} \Big) \nonumber\\
	&\,\qquad\qquad\qquad \cdot\Big( \boldsymbol\epsilon_{\mbf k,\lambda} \hat{a}_{\mbf k,\lambda} e^{i(\mbf k\mbf r-\omega_kt)} + \text{h.c.} \Big)
\end{align}
with $\omega_{mn}\equiv \omega_{e_m}-\omega_{g_n}$.

%%% ----------------------------------------------------------------------------
%%%				Subsection
%%% ----------------------------------------------------------------------------

\subsection{Born-Markov approximation}

We perform the usual Born-Markov approximation by assuming that (i) the atoms and the field degrees of freedom are essentially uncorrelated (i.e. very short correlation time), and (ii) the state of the field is independent of time. For this we consider a window of time $\Delta t=t-t_0$ small enough such that the state of the atoms does not change, but much larger than the correlation time between atoms and field. Integrating Eq.~(\ref{eq:intpic_schroed}) from $t$ to $t_0$ we get
\begin{equation}
	\hat{\rho}_{af}(t) = \hat{\rho}_{af}(t_0) +\frac{1}{i\hbar} \int_0^{\Delta t} d\tau \left[ \hat{V}(t-\tau) , \hat{\rho}_{af}(t-\tau) \right].
\end{equation}
Inserting this again into Eq.~(\ref{eq:intpic_schroed}), using $\hat{\rho}_{af}(t-\tau)\approx \hat{\rho}_{a}(t-\tau)\otimes \hat{\rho}_f$, $\hat{\rho}_{a}(t-\tau)\approx \hat{\rho}_{a}(t_0) \approx \hat{\rho}_{a}(t)$, and tracing out over the field degrees of freedom, we obtain the master equation
\begin{align}
	\frac{d\hat{\rho}_a}{dt} =&\, \frac{1}{i\hbar} \Tr_f \left[ \hat{V}(t), \hat{\rho}_a(t)\otimes\hat{\rho}_f \right] \nonumber\\
	&\,- \frac{1}{\hbar^2} \Tr_f \int_0^{\Delta t} d\tau \left[ \hat{V}(t) , \left[ \hat{V}(t-\tau) , \hat{\rho}_a(t)\otimes\hat{\rho}_f \right] \right] \nonumber \\[5pt]
	\equiv&\, \Delta\hat{\rho}^{(1)} + \Delta\hat{\rho}^{(2)} .
\end{align}
In the presence of a coherent laser field, the first-order term $\Delta\hat{\rho}^{(1)}$ gives rise to the usual Rabi Hamiltonian, which is derived further below. Here, we consider first the case of vacuum electromagnetic field, for which the first-order term vanishes. In this case, the second-order term gives rise to dipolar interactions, as we show next.
In the following, we drop the subscript $a$ and write $\hat\rho$ for the atomic density matrix.

For vacuum background field only the $\hat{a}^\dagger \hat{a}$ and $\hat{a} \hat{a}^\dagger$ terms in $\Delta\hat{\rho}^{(2)}$ with $\mbf k=\mbf k'$ and $\lambda=\lambda'$ survive.
Assuming the energy differences $\omega_{e_m}-\omega_{g_n}$ are much larger than the energy differences within the ground/excited manifolds (which is true for optical frequencies), we further neglect all terms of the form $\hat\sigma \hat\sigma$ and $\hat\sigma^\dagger \hat\sigma^\dagger$.
Going back to the frame rotating with $\hat{H}_\text{atom}$ as $\hat\rho\equiv e^{-i\hat{H}_\text{atom} t/\hbar}\hat\rho_a\,e^{i\hat{H}_\text{atom} t/\hbar}$ the second-order term then reads
\begin{align}
	\Delta\hat\rho^{(2)} =&\, -\frac{1}{\hbar^2} \int_{\mbf r,\mbf r'} \sum_{m,n} \sum_{m',n'} \sum_{\mbf k,\lambda} \mathrm{g}_k^2 \int_0^{\Delta t} d\tau \left( \hat{T}_1 + \hat{T}_2 + \text{h.c.} \right) ,
\label{eq:quadr_part_gen}
\end{align}
with
\begin{align}
	\hat{T}_1 \equiv&\, e^{i\mbf k \Delta\mbf r} e^{-i\tau(\omega_k+\omega_{m'n'})} P_1\, \Sigma_1 ,\\
	\hat{T}_2 \equiv&\, e^{i\mbf k \Delta\mbf r} e^{-i\tau(\omega_k-\omega_{m'n'})} P_2\, \Sigma_2,
\end{align}
where $\Delta\mbf r\equiv\mbf r-\mbf r'$ and
\begin{align}
	\hat{\Sigma}_1 \equiv&\, \hat{\sigma}^{\dagger}_{e_mg_n}(\mbf r)\, \hat{\sigma}_{e_{m'}g_{n'}}(\mbf r')\, \hat{\rho} - \hat{\sigma}_{e_{m'}g_{n'}}(\mbf r')\, \hat{\rho}\, \hat{\sigma}^{\dagger}_{e_mg_n}(\mbf r), \\
	\hat{\Sigma}_2 \equiv&\, \hat{\sigma}_{e_mg_n}(\mbf r)\, \hat{\sigma}^{\dagger}_{e_{m'}g_{n'}}(\mbf r')\, \hat{\rho} - \hat{\sigma}^{\dagger}_{e_{m'}g_{n'}}(\mbf r')\, \hat{\rho}\, \hat{\sigma}_{e_mg_n}(\mbf r), \\
	P_1 \equiv&\, \big( \mbf d_{e_mg_n}^*\cdot \boldsymbol\epsilon_{\mbf k,\lambda} \big) \big( \mbf d_{e_{m'}g_{n'}}\cdot \boldsymbol\epsilon_{\mbf k,\lambda} \big), \\
	P_2 \equiv&\, \big( \mbf d_{e_mg_n} \cdot \boldsymbol\epsilon_{\mbf k,\lambda} \big) \big( \mbf d_{e_{m'}g_{n'}}^* \cdot \boldsymbol\epsilon_{\mbf k,\lambda} \big).
\end{align}

Next, we perform the integral over time and the sum over the momentum modes and polarization directions.
For the time integral we make the approximation
\begin{equation}
	\int_0^{\Delta t} d\tau \quad\longrightarrow\quad \int_0^\infty d\tau .
\end{equation}
Using the well-known identity
\begin{equation}
	\int_0^\infty c\, d\tau\, e^{ic(k\pm k_0)\tau} = \pi\,\delta(k\pm k_0) + i \mathcal{P} \frac{1}{k\pm k_0} ,
\end{equation}
where $\mathcal{P}$ denotes the principal value, the time integral in Eq.~(\ref{eq:quadr_part_gen}) yields
\begin{align}
	&\,\int_0^\infty c d\tau (\hat{T}_1+\hat{T}_2)\\
	&\,= e^{i\mbf k \Delta\mbf r} \left\{ P_1\, \hat{\Sigma}_1 \left( \pi\,\delta(k+k_{m'n'}) - i\mathcal{P} \frac{1}{k+k_{m'n'}} \right) \right. \nonumber\\
	&\,\qquad\quad + \left. P_2\, \hat{\Sigma}_2 \left( \pi\,\delta(k-k_{m'n'}) - i\mathcal{P} \frac{1}{k-k_{m'n'}} \right) \right\} ,
\end{align}
where we defined $k_{mn} \equiv \omega_{mn}/c$.

Using the fact that the two polarizations $\boldsymbol\epsilon_{\mbf k,\lambda}$ and $\hat{\mbf k}\equiv \mbf k/|\mbf k|$ from an orthonormal system, we can perform the sum over polarization directions, $\sum_\lambda P_{1,2}$, as
\begin{align}
	\sum_\lambda \big( \mbf d_1 \cdot \boldsymbol\epsilon_{\mbf k,\lambda} \big) \big( \mbf d_2 \cdot \boldsymbol\epsilon_{\mbf k,\lambda} \big) =&\, \mbf d_1\cdot\mbf d_2- \big( \hat{\mbf k}\cdot \mbf d_1 \big) \big( \hat{\mbf k}\cdot \mbf d_2 \big) \nonumber\\[-5pt]
	=&\, \mbf d_1^T \big( \mathbb{1} - \hat{\mbf k} \otimes \hat{\mbf k} \big)\, \mbf d_2 \nonumber\\
	\equiv&\, \mbf d_1^T P(\hat{\mbf k})\, \mbf d_2
\end{align}
where $\mbf d_{1,2}$ stands for dipole matrix elements, e.g.~$\mbf d_{e_mg_n}$. Here, $\mathbb{1}$ is a 3-dimensional unit matrix and $(\mbf a\otimes\mbf b)_{\alpha\beta}=a_\alpha b_\beta$.

To perform the sum over the momentum modes $\mbf k$ we take the continuum limit as
\begin{equation}
	\sum_{\mbf k} \quad\longrightarrow\quad \frac{V}{(2\pi)^3} \int d^3k .
\end{equation}
We start with the angular part of the $\mbf k$-integral,
\begin{equation}
	I_\Omega \equiv \int_0^{2\pi} d\varphi_k \int_0^\pi d\theta_k\, \sin\theta_k\, e^{i\mbf k \Delta\mbf r} P(\hat{\mbf k}).
\end{equation}
This can be computed by expressing the vector $\mbf k$ in a coordinate system with $\Delta\mbf r$ as $z$-axis. In this way, one obtains $I_\Omega=\tilde{G}_I(k,\Delta\mbf r)$ with
\begin{align}
	\tilde{G}_I (k,\mbf r) =&\, 4\pi \left\{ \mathbb{1} \left[ \frac{\sin(kr)}{kr} + \frac{\cos(kr)}{(kr)^2} - \frac{\sin(kr)}{(kr)^3} \right] \right. \nonumber\\
	&\, \left. + \hat{\mbf r} \otimes \hat{\mbf r} \left[ - \frac{\sin(kr)}{kr} - 3 \frac{\cos(kr)}{(kr)^2} + 3 \frac{\sin(kr)}{(kr)^3} \right] \right\} ,
\end{align}
which is related to the electromagnetic Green tensor in vacuum.
In the limit $\Delta\mbf r\rightarrow 0$ one can show that $I_\Omega = \frac{8\pi}{3} \mathbb{1}$.

Putting everything so far together, the quadratic contribution to the master equation is given by
\begin{equation}
	\Delta\hat{\rho}^{(2)} = - \frac{1}{\hbar^2} \int_{\mbf r,\mbf r'} \sum_{m,n} \sum_{m',n'} ( I_k + \text{h.c.} )
\end{equation}
with
\begin{align}
	I_k \equiv&\, \frac{\hbar}{2 \varepsilon_0 (2\pi)^3} \int_0^\infty dk\, k^3 \nonumber\\
	&\, \left\{ \tilde{P}_1\, \Sigma_1 \left( \pi\,\delta(k+k_{m'n'}) - i\mathcal{P} \frac{1}{k+k_{m'n'}} \right) \right. \nonumber\\
	&\, \left. + \tilde{P}_2\, \Sigma_2 \left( \pi\,\delta(k-k_{m'n'}) - i\mathcal{P} \frac{1}{k-k_{m'n'}} \right) \right\},
\label{eq:Ik_master}
\end{align}
and $\tilde{P}_1\equiv\mbf d_{e_mg_n}^{*T}\, \tilde{G}_I(k,\Delta\mbf r)\, \mbf d_{e_{m'}g_{n'}}$, $\tilde{P}_2\equiv\tilde{P}_1^*$.
The $\delta$-functions in this expression give rise to the incoherent Linbladian part of the master equation, whereas the principal value part leads to the coherent Hamiltonian part.
We write $\Delta\hat{\rho}^{(2)}\equiv \Delta\hat{\rho}^{(2)}_\text{coh} + \Delta\hat{\rho}^{(2)}_\text{inc}$ and give the two contributions separately in the following.

Before doing so, we do an additional approximation by assuming that $|k_{mn}-k_{m'n'}|\ll (k_{mn}+k_{m'n'})$, i.e.~the typical energy differences between different ground (excited) states is small compared to the gap between the ground and excited manifolds (this is justified for optical transitions). This allows to substitute $k_{m'n'}\approx k_{mn}\rightarrow k_0$ in the expressions above, where $k_0$ can be seen as the mean frequency difference between the $e$ and $g$ manifolds.
One can imagine this as a Taylor expansion around $k_0$ up to leading non-vanishing order.

Since $k_{m'n'}>0$ the $\delta(k+k_{m'n'})$ terms vanish and the incoherent contribution becomes
\begin{align}
	&\Delta\rho^{(2)}_{\text{inc}} = - \int_{\mbf r,\mbf r'} \sum_{m,n} \sum_{m',n'} \big(\mbf d_{mn}^{T}\, \Im G(k_0,\mbf r-\mbf r')\, \mbf d_{m'n'}^*\big) \nonumber\\
	&\,\quad \left( \left\{ \hat{\sigma}_{e_mg_n}(\mbf r)\, \hat{\sigma}^{\dagger}_{e_{m'}g_{n'}}(\mbf r'), \hat{\rho} \right\} - 2 \hat{\sigma}^{\dagger}_{e_{m'}g_{n'}}(\mbf r')\, \hat{\rho}\, \hat{\sigma}_{e_mg_n}(\mbf r) \right),
\label{eq:drho2_inc}
\end{align}
where the electromagnetic tensor $G$ is defined in Eq.~(\ref{eq:Gtensor}), and we now included the $k_0$ dependence explicitly.
For $\mbf r\rightarrow0$ one obtains $\Im G(\mbf r) \rightarrow \frac{\Gamma}{2}\mathbb{1}$.
In this expression, we used the Wigner-Eckart decomposition of the dipole matrix elements given in the main text. The radial part has been absorbed into $\Gamma$, and we dropped the superscript `sph' from the spherical part, $\mbf d_{mn}\equiv \mbf d_{mn}^\text{sph}$.

After combining the terms in Eq.~(\ref{eq:Ik_master}) with their hermitian conjugates, the terms proportional to $\hat{\sigma}^\dagger \hat{\rho} \hat{\sigma}$ and $\hat{\sigma} \hat{\rho} \hat{\sigma}^\dagger$ vanish and the rest reduces after some reordering to
\begin{align}
	I_k^{\text{coh}} =&\, \frac{-i\hbar}{4\pi^2\varepsilon_0} \mbf d_{e_mg_n}^{T} \left( \mathcal{P} \int_{-\infty}^\infty dk \frac{k^3 \Im G(k,\mbf r-\mbf r')}{k-k_0} \right) \mbf d_{e_{m'}g_{n'}}^*\nonumber\\
	&\, \times \hat{\sigma}_{mn}(\mbf r)\, \hat{\sigma}^{\dagger}_{m'n'}(\mbf r')\, \hat{\rho} .
\end{align}
Using contour integration the coherent contribution to the master equation becomes
\begin{align}
	\Delta\hat{\rho}^{(2)}_{\text{coh}} \equiv&\, - \frac{1}{i} \int_{\mbf r,\mbf r'} \sum_{m,n} \sum_{m',n'} \big(\mbf d_{mn}^{T}\, \Re G(k_0,\mbf r-\mbf r')\, \mbf d_{m'n'}^*\big)\nonumber\\
	&\, \qquad\qquad\quad \Big[ \sigma_{e_mg_n}(\mbf r)\, \sigma^{\dagger}_{e_{m'}g_{n'}}(\mbf r') , \hat{\rho} \Big] .
\label{eq:drho2_coh}
\end{align}
Strictly speaking, $\Re G(k_0,\mbf r-\mbf r')$ diverges as $\Delta\mbf r\rightarrow0$, but this will be smoothen out once the integrals over $\mbf r$ and $\mbf r'$ are carried out.

Equations (\ref{eq:drho2_coh}) and (\ref{eq:drho2_inc}) are already very close to the final expression for the master equation given in the main text. To finish the derivation we just need to expand the operators in the basis of Wannier functions and write the dipole matrix elements in terms of Clebsch-Gordan coefficients.

As a side remark, we notice that in the limit where the energy difference $\omega_{mn}-\omega_{m'n'}$ for some $(m,n)\neq(m',n')$ is much larger than $\Gamma$, the corresponding terms $\hat\sigma_{e_mg_n}\sigma^\dagger_{e_{m'}g_{n'}}$ are fast-rotating and can be ignored.

%%% ----------------------------------------------------------------------------
%%%				Subsection
%%% ----------------------------------------------------------------------------

\subsection{Wannier basis\label{app:wannier}}

We assume the atoms are resting in the lowest band of an optical lattice and expand the field operators in the Wannier basis as $\hat\psi_{a_m}(\mbf r) = \sum_i w_i(\mbf r)\, \hat c_{i,a_m}$, where $a\in\{g,e\}$ and $i$ denotes lattice site. We make a harmonic approximation and set $w_i(\mbf r) \approx \psi_0(\mbf r-\mbf r_i)$, where $\psi_0(\mbf r) = \prod_{n=\{x,y,z\}} \frac{1}{(2\pi \ell_n^2)^{1/4}} e^{-r_n^2/(4\ell_n^2)}$ is the ground state of the oscillator, $\ell_n^2 \equiv \hbar/(2m\omega_n)$, $\omega_n \equiv (2\,V_n\, k_L^2/m)^{1/2}$ are the onsite trap frequencies, $k_L$ is the trap wave-vector, $m$ is the mass, and $V_n$ is the trap depth in the corresponding direction, $V(\mbf r)=\sum_{n=\{x,y,z\}} V_n \sin(k_Lr_n)^2$.

We insert the Wannier expansion into the above expressions and obtain in this way $d\hat{\rho}/dt=-i[\hat{H},\hat{\rho}]+\mathcal{L}(\hat{\rho})$ with
\begin{align}
	\hat{H} =&\, - \sum_{m,n} \sum_{m'n'} \sum_{ij} \big(\mbf d_{mn}^{T}\, \Re G^{ij}\, \mbf d_{m'n'}^*\big)\, \hat\sigma^{(i)}_{e_mg_n}\, \hat\sigma^{(j)}_{g_{n'}e_{m'}} ,
\label{eq:Hdip_dsigma}\\
	\mathcal{L}(\hat{\rho}) =&\, - \sum_{m,n} \sum_{m',n'} \sum_{ij} \big(\mbf d_{mn}^{T}\, \Im G^{ij}\, \mbf d_{m'n'}^*\big)\nonumber\\
	&\, \left( \left\{ \hat\sigma^{(i)}_{e_mg_n}\, \hat\sigma^{(j)}_{g_{n'}e_{m'}}, \hat{\rho} \right\} - 2 \hat\sigma^{(j)}_{g_{n'}e_{m'}}\, \hat{\rho}\, \hat\sigma^{(i)}_{e_mg_n} \right) ,
\label{eq:Ldip_dsigma}
\end{align}
where we defined $\hat\sigma^{(i)}_{g_me_n} \equiv \hat c^\dagger_{i,g_m} \hat c_{i,e_n}$ as in the main text.
The interaction tensors are given by
\begin{align}
	\Im G^{ij} \equiv&\, \int d\mbf r\,d\mbf r'\, \Im G(k_0,\mbf r-\mbf r')\, |w_i(\mbf r)|^2\, |w_j(\mbf r')|^2,
\label{eq:fij_def}\\
	\Re G^{ij} \equiv&\, \int d\mbf r\,d\mbf r'\, \Re G(k_0,\mbf r-\mbf r')\, |w_i(\mbf r)|^2\, |w_j(\mbf r')|^2.
\label{eq:gij_def}
\end{align}
For $i\neq j$ we approximate the square of the Wannier functions by $\delta$-functions and obtain $\Im G^{ij}\approx \Im G(k_0,\mbf r_i-\mbf r_j)$ and $\Re G^{ij}\approx \Re G(k_0,\mbf r_i-\mbf r_j)$. The on-site case $i=j$ requires a more careful treatment and will be computed in App.~\ref{app:onsite}.

Using the expression of the dipole matrix elements in terms of Clebsch-Gordan coefficients, Eq.~(\ref{eq:dsph_clebsch}), and the definition of the multilevel raising/lowering operators, Eq.~(\ref{eq:D-def}), the Hamiltonian and Linbladian of Eqs.~(\ref{eq:Hdip_dsigma}) and (\ref{eq:Ldip_dsigma}) turn into the expressions given in the main text, Eq.~(\ref{eq:HLdipoles}).

%%% ----------------------------------------------------------------------------
%%%				Subsection
%%% ----------------------------------------------------------------------------

\subsection{Single-particle Hamiltonian}

We give here the single-body Hamiltonian missing from the master equation presented above, which we write as
\begin{align}
	\hat{H}_\text{single} = \hat{H}_L + \hat{H}_\Delta.
\end{align}
We assume the atoms are subject to a laser with electric field given by $\mbf E (t,\mbf r) = \boldsymbol\epsilon_L \mathcal{E} e^{i(\mbf k_L\mbf r - \omega_{L} t)} + \text{c.c.}$ with amplitude $\mathcal{E}$, polarization $\boldsymbol\epsilon_L$, frequency $\omega_L$ and wave-vector $\mbf k_L$.
The first-order term of the master equation is then given by $\Delta\hat{\rho}^{(1)} = 1/(i\hbar) [ \hat{H}_L , \hat{\rho} ]$, with
\begin{align}
	\hat{H}_L =&\, - \hbar \sum_{i,m,n} \left[ \Omega_{e_mg_n}\, \hat{\sigma}^{(i)}_{e_mg_n}\, e^{i\mbf{k}_L \mbf r_i} + \text{h.c.} \right] ,
\end{align}
and $\Omega_{e_mg_n} \equiv ( \mbf d_{e_mg_n} \cdot \boldsymbol\epsilon_L ) \mathcal{E}/\hbar\equiv \Omega ( \mbf d^\text{sph}_{mn} \cdot \boldsymbol\epsilon_L ) $. Here, we went into the rotating frame of the laser using $\hat{\rho}\rightarrow e^{i \hat{H}_{\omega_L} t/\hbar} \hat{\rho} e^{-i \hat{H}_{\omega_L} t/\hbar}$ with $\hat{H}_{\omega_L}=\sum_{i,m} \hbar\omega_L\hat{\sigma}^{(i)}_{e_me_m}$ and performed a Rotating Wave Approximation. Thus, the Hamiltonian for the level energies is given by
\begin{align}
	\hat{H}_\Delta =&\, \hbar\sum_{i,m} (\omega_{e_m}-\omega_L) \hat\sigma^{(i)}_{e_me_m} + \hbar\sum_{i,n} \omega_{g_n} \hat\sigma^{(i)}_{g_ng_n}.
\end{align}

%%%%%%%%%%%%%					%%%%%%%%%%%%%
%%%%%%%%%%%%%		SECTION		%%%%%%%%%%%%%
%%%%%%%%%%%%%					%%%%%%%%%%%%%

\section{Onsite interactions for a deep trap\label{app:onsite}}

As shown in App.~\ref{app:master_eq_deriv}, the onsite dipole interaction matrices are given by
\begin{align}
	G^{ii} = \int d\mbf r\, d\mbf r'\, G(\mbf r-\mbf r')\, |w_i(\mbf r)|^2\, |w_i(\mbf r')|^2,
\end{align}
where $G(\mbf r-\mbf r')$ is the dyadic Green's tensor defined in the main text, and $w_i(\mbf r)=w(\mbf r-\mbf r_i)$ is the Wannier function of the lowest motional eigenstate of lattice site $i$.
We consider here the limit of a deep trap, such that $w_i(\mbf r)\approx \psi_0(\mbf r-\mbf r_i)$ can be approximated by the ground-state wave-function of a harmonic potential, $\psi_0(\mbf r) = \prod_{n=\{x,y,z\}} \frac{1}{(2\pi \ell_n^2)^{1/4}} e^{-r_n^2/(4\ell_n^2)}$.
Note that the components $r_n$ of the position vector $\mbf r$ are given with respect to the coordinate system defined by the lattice axes.

The leading-order terms of $G(\mbf r)$ for short distances are given by
\begin{align}
	\Re G(\mbf r) \approx&\,  -\frac{3\Gamma}{4} \frac{ \mathbb{1}-3\hat{r} \otimes \hat{r} }{ (k_0r)^3 } ,
\label{eq:ReG_onsite_approx}\\
	\Im G(\mbf r) \approx&\, \frac{\Gamma}{2} \mathbb{1}.
\label{eq:ImG_onsite_approx}
\end{align}
This implies $\Im G^{ii}=\frac{\Gamma}{2}\mathbb{1}$.
In the following, we explicitly compute the integral $\Re G^{ii}$ using the approximation of Eq.~(\ref{eq:ReG_onsite_approx}).

%%% ----------------------------------------------------------------------------
%%%				Subsection
%%% ----------------------------------------------------------------------------

\subsection{Onsite coefficient}

The aim is to compute the integral
\begin{align}
	\Re G^{ii} = -\frac{3\Gamma}{4k_0^3} \int d\mbf r\, d\mbf r'\, \frac{ \mathbb{1}-3\,\Delta\hat{\mbf r} \otimes \Delta\hat{\mbf r} }{ |\Delta\mbf r|^3 }\, \rho_0(\mbf r)\, \rho_0(\mbf r'),
\end{align}
which is independent of $i$, and where we defined $\rho_0(\mbf r)\equiv|\psi_0(\mbf r)|^2$, $\Delta\mbf r=\mbf r-\mbf r'$ and $\Delta\hat{\mbf r} = \Delta\mbf r/|\Delta\mbf r|$.
We start by expanding the vector $\Delta\hat{\mbf r}$ in spherical components with respect to the lattice axes $\{\mbf e^L_x,\mbf e^L_y,\mbf e^L_z\}$ as
\begin{equation}
	\Delta\hat{\mbf r} = -\frac{1}{\sqrt{2}} \sin\theta\, e^{i\varphi} \mbf e^L_+ + \frac{1}{\sqrt{2}} \sin\theta\, e^{-i\varphi} \mbf e^L_- + \cos\theta\, \mbf e^L_z,
\end{equation}
where $\mbf e^L_\pm = \mp(\mbf e^L_x\pm i\mbf e_y^L)$ as usual.
Note that the prefactors are proportional to the spherical harmonics $Y_{l,m}$ for $l=1$. Using this expansion we can write
\begin{align}
	&\Re G^{ii} =  -\frac{3\Gamma}{4k_0^3}\times \nonumber\\
	 &\,\bigg\{ V^{0} \left[ \frac{1}{2}\left( \mbf e^L_+\otimes\mbf e^L_- + \mbf e^L_-\otimes\mbf e^L_+ \right) + \mbf e^L_z\otimes\mbf e^L_z \right]\nonumber\\
	&\,+ V^{+z} \left( \mbf e^L_+\otimes\mbf e^L_z + \mbf e^L_z\otimes\mbf e^L_+ \right) - V^{-z} \left( \mbf e^L_-\otimes\mbf e^L_z + \mbf e^L_z\otimes\mbf e^L_- \right) \nonumber\\
	&\,+ V^{++} \mbf e^L_+\otimes\mbf e^L_+ + V^{--} \mbf e^L_-\otimes\mbf e^L_- \bigg\}
\label{eq:dipole_expansion1}
\end{align}
where we defined the onsite coefficients
\begin{align}
	V^{0} \equiv&\, \int d\mbf r\,d\mbf r'\, V^{0}(\mbf r-\mbf r')\, \rho_0(\mbf r)\, \rho_0(\mbf r'), 
\label{eq:V0_def}\\
	V^{\pm z} \equiv&\, \int d\mbf r\,d\mbf r'\, V^{\pm z}(\mbf r-\mbf r')\, \rho_0(\mbf r)\, \rho_0(\mbf r'), 
\label{eq:Vpmz_def}\\
	V^{\pm\pm} \equiv&\,  \int d\mbf r\,d\mbf r'\, V^{\pm\pm}(\mbf r-\mbf r') \, \rho_0(\mbf r)\, \rho_0(\mbf r'),
\label{eq:Vpmpm_def}
\end{align}
with
\begin{align}
	V^{0}(\mbf r) \equiv&\, \frac{1-3\cos\theta^2}{r^3} ,
\label{eq:V0r_def}\\
	V^{\pm z}(\mbf r) \equiv&\, \frac{3}{\sqrt{2}} \frac{\sin\theta\cos\theta e^{\pm i\varphi}}{r^3},
\label{eq:Vpmzr_def}\\
	V^{\pm\pm}(\mbf r) \equiv&\, -\frac{3}{2} \frac{\sin\theta^2 e^{\pm 2i\varphi}}{r^3}.
\label{eq:Vpmpmr_def}
\end{align}
The problem is thus reduced to the computation of the coefficients $V^0$, $V^{\pm z}$ and $V^{\pm\pm}$.

All these expressions have the form of convolution integrals, $V^{\times}=\int d\mbf r\,d\mbf r'\, V^\times(\mbf r-\mbf r')\, \rho_0(\mbf r)\, \rho_0(\mbf r')$, where `$\times$' stands for any of the above superscripts. Using the convolution theorem, one can compute them as
\begin{equation}
	V^{\times} =  \int \frac{d\mbf k}{(2\pi)^3}\, V^\times(\mbf k)\, \rho_0(\mbf k)\, \rho_0(-\mbf k),
\label{eq:Vconvolution}
\end{equation}
where the Fourier transform is defined as $ V^\times(\mbf k) = \int d\mbf r\, e^{-i\mbf k\mbf r}\, V^\times(\mbf r)$, and equivalently for $\rho_0$.
The Fourier transformed functions can be explicitly computed to be
\begin{align}
	V^{0}(\mbf k) =&\, -\frac{4\pi}{3} \big( 1 - 3\cos\theta_k^2 \big),
\label{eq:V0_fourier}\\
	V^{\pm z}(\mbf k) =&\, -\frac{4\pi}{3} \left( \frac{3}{\sqrt{2}} \sin(\theta_k)\,\cos( \theta_k) \, e^{\pm i\varphi_k} \right),
\label{eq:Vpmz_fourier}\\
	V^{\pm\pm}(\mbf k) =&\, -\frac{4\pi}{3} \left( -\frac{3}{2} \sin\theta_k^2\, e^{\pm2i\varphi_k} \right) ,
\label{eq:Vpmpm_fourier}\\
	\rho_0(\mbf k) =&\, e^{-\frac{1}{2}\sum_n k_n^2 \ell_n^2} = \rho_0(-\mbf k).
\end{align}
where $\varphi_k$ and $\theta_k$ are the spherical angles of $\mbf k$ in the lattice coordinate system.
Up to a prefactor $-4\pi/3$, the functions $V^\times(\mbf k)$ have all the same form as Eqs.~(\ref{eq:V0r_def}), (\ref{eq:Vpmzr_def}), and (\ref{eq:Vpmpmr_def}) without the $1/r^3$. Therefore, the remaining integral can be written as
\begin{align}
	\Re G^{ii} = \frac{\Gamma\pi}{k_0^3} \int \frac{d\mbf k}{(2\pi)^3}\, (\mathbb{1}-3\,\hat{\mbf k} \otimes \hat{\mbf k} )\, e^{-\sum_n k_n^2\, \ell_n^2},
\end{align}
In principle, the integral over $\mbf k$ can now be computed numerically for given values of $\ell_{x,y,z}$.

We consider an axially symmetric trap, $\ell_x=\ell_y\equiv\ell_\perp$. Performing the integrals over $\varphi_k$ and $k$ yields
\begin{align}
	\Re G^{ii} = \frac{3\Gamma}{4} U(k_0,\ell_z,\ell_\perp) (\mathbb{1}-3\,\mbf e^L_z \otimes \mbf e^L_z ),
\end{align}
with the prefactor
\begin{equation}
	U(k_0,\ell_z,\ell_\perp) = \frac{1}{24\sqrt{\pi}k_0^3} \int_{-1}^1 dx\, \frac{3x^2-1}{[(\ell_z^2-\ell_\perp^2)x^2+\ell_\perp^2]^{3/2}}.
\label{eq:onsite_integral}
\end{equation}
Recall that $\mbf e^L_z$ is the $z$-axis of the lattice and is not necessarily the same as the quantization axis $\mbf e_z$.

%%% ----------------------------------------------------------------------------
%%%				Subsection
%%% ----------------------------------------------------------------------------

\subsection{Scaling analysis of onsite coefficient}

We analyze now the dependence of the onsite interaction prefactor (\ref{eq:onsite_integral}) on the laser wave-length $\lambda_0$, lattice wave-length $\lambda_L$ and lattice depth. The widths $\ell_n$ of the ground-state wave function $\psi_0(\mbf r)$ are given by
\begin{equation}
	\ell_n^2 = \frac{\lambda_L^2}{8\pi^2 \sqrt{\nu_n}},
\label{eq:sigma}
\end{equation}
where $\nu_n=V_n/E_R$ is the depth of the lattice potential in units of the photon recoil energy $E_R=\hbar^2k_L^2/(2m)$, see App.~\ref{app:wannier}.

The onsite prefactor scales as
\begin{equation}
	U(k_0,\ell_z,\ell_\perp) = \left( \frac{\lambda_0}{\lambda_L} \right)^3\, (\nu_z\nu_\perp)^{3/8}\, \tilde{U}(\ell_z/\ell_\perp),
\end{equation}
where $\tilde{U}$ is proportional to the integral in Eq.~(\ref{eq:onsite_integral}) and needs to be computed numerically.
This expression implies that for fixed ratio $\ell_z/\ell_\perp$ we can make the onsite prefactor $U$ larger by increasing the lattice depth, although the increase is relatively slow with $(\nu_z\nu_\perp)^{3/8}$. Alternatively, one may use a different lattice laser or transition frequency, so as to make $\lambda_0/\lambda_L$ larger.

\begin{figure}[t!]
\centering
	\includegraphics[width=\columnwidth]{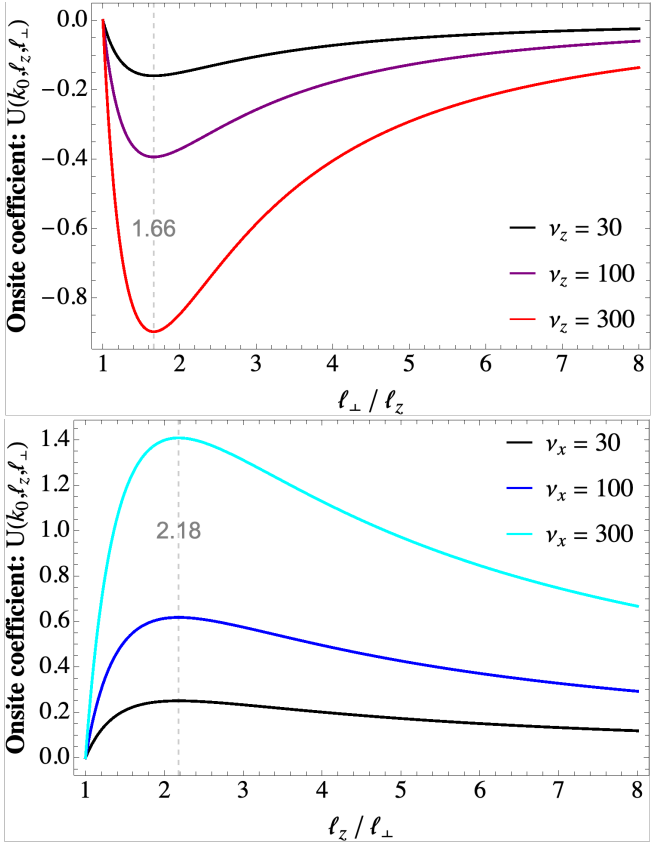}
	\caption{\emph{Top:} Onsite prefactor $U(k_0,\ell_z,\ell_\perp)$ as a function of $\ell_\perp/\ell_z$. \emph{Bottom:} Onsite prefactor $U(k_0,\ell_z,\ell_\perp)$ as a function of $\ell_z/\ell_\perp$. Both are plotted for different trap depths $\nu_n=V_n/E_R$ in units of the recoil energy, using $\lambda_0=689\,\text{nm}$ and $\lambda_L=813\,\text{nm}$.}
	\label{fig:selfI}
\end{figure}

Figure~\ref{fig:selfI} shows plots of $U$ for fixed $\ell_z$ ($\ell_\perp$) as a function of $\ell_\perp/\ell_z$ ($\ell_z/\ell_\perp$) for different lattice depths. For this we used $\lambda_0=689\,\text{nm}$ and $\lambda_L=813\,\text{nm}$, which correspond to the $^1\text{S}_0\rightarrow ^3\text{P}_0$ transition of $^{87}\text{Sr}$ on a magic wave-length lattice~\cite{LudlowRevModPhys87}. Interestingly, in both cases the function reaches a maximum absolute value at a given ratio and then decreases monotonously. This can be understood from the fact that when the ratio $\ell_\perp/\ell_z$ ($\ell_z/\ell_\perp$) approaches 1, the integral vanishes because of symmetry. On the other hand, when the ratio is very large the interaction approaches zero because the atoms are on average too far apart.

We numerically estimate the maximum to be at
\begin{align}
	\argmax_{\ell_\perp/\ell_z>1} \left| U(\ell_\perp/\ell_z)\big|_{\ell_z=\text{const}} \right| \approx&\, 1.66,\\
	\argmax_{\ell_z/\ell_\perp>1} | U(\ell_z/\ell_\perp)\big|_{\ell_\perp=\text{const}} | \approx&\, 2.18.
\end{align}
Using $\ell_\perp/\ell_z=(\nu_z/\nu_\perp)^{1/4}$ this implies that the ratio of lattice depths at which the onsite interaction is maximal is $\nu_z/\nu_\perp\approx7.59$ and $\nu_\perp/\nu_z\approx22.59$, respectively. The maximum value achieved is larger for the case $\ell_z/\ell_\perp>1$, i.e.~for cigar-shaped rather than pancake-shaped traps.

%%%%%%%%%%%%%					%%%%%%%%%%%%%
%%%%%%%%%%%%%		SECTION		%%%%%%%%%%%%%
%%%%%%%%%%%%%					%%%%%%%%%%%%%

\section{Dipole operator\label{app:dipole_op}}

The dipole operator $\hat{\mbf d}$ can be expanded (ignoring the site index $i$) as
\begin{align}
	\hat{\mbf d} =&\, \sum_{m,n}  \big(\mbf d_{e_mg_n} \hat{\sigma}_{e_mg_n} + \text{h.c.}  \big),
\end{align}
which was used in Eq.~(\ref{eq:Hatomfield}). The dipole operator is a vector under rotations and thus can be expanded in spherical components as $\hat{\mbf d}=\sum_q (-1)^q \mbf e_{-q} \hat{d}^1_q$, where $\hat{d}^1_0 = \hat{d}_z$, $\hat{d}^1_{\pm} = \mp\big( \hat{d}_x \pm i\hat{d}_y \big)/\sqrt{2}$. Using the Wigner-Eckart theorem and the symmetries of the Clebsch-Gordan coefficients~\cite{brown2003rotational}, one can then show
\begin{align}
	\mbf d_{e_mg_n} =&\, \sum_q (-1)^q \mbf e_{-q} \langle e_m | \hat{d}^1_q | g_n \rangle \nonumber\\
	=&\, \frac{\langle e,f_e || \hat{d}^1 || g,f_g\rangle}{\sqrt{2f_e+1}} \sum_q (-1)^q \mbf e_{-q} C^{q}_{n}.
\end{align}
with $C^{q}_{n}$ as defined in Sec.~\ref{ssec:light_matterH}, and the reduced dipole matrix element $d^\text{rad}_{ge}=\langle e,f_e || \hat{d}^1 || g,f_g\rangle$.
Inserting this into the above expression and reordering terms one can then split the dipole operator into
\begin{align}
	\hat{\mbf d} =&\, \frac{\langle e,f_e || \hat{d}^1 || g,f_g\rangle}{\sqrt{2f_e+1}} \left(\Dv^+ + \Dv^- \right)
\end{align}
where
\begin{align}
	\Dv^+ \equiv&\, \sum_q (-1)^q \mbf e_{-q} \D^+_{q}, \\
	\Dv^- \equiv&\, \sum_q (-1)^q \mbf e_{-q}^* \D^-_{q}.
\end{align}
Since the dipole operator is a vector, and the operators $\Dv^\pm$ do not mix under rotations, it follows that the three components ($q=0,\pm1$) of $\Dv^+$ ($\Dv^-$) also form a vector. We note, however, that given the above expressions, the spherical components of $\Dv^+$ are $\D^+_q$, whereas the spherical components of $\Dv^-$ are strictly speaking $(-1)^q\D^-_{-q}$. This detail should be taken into account when explicitly evaluating matrix elements.

To compute matrix elements of the operators $\D^\pm_q$ with respect to eigenstates of the total angular momentum we use the fact that they are spherical operators. Thus, using the Wigner-Eckart theorem it follows as above that
\begin{align}
	\langle \eta_1, f_1,m_1 | \D^+_q | \eta_2,f_2,m_2\rangle &\, \nonumber\\
	= \frac{\langle \eta_1,f_1 || \D^+ || \eta_2,f_2 \rangle}{\sqrt{2f_1+1}}&\, \langle f_2\,m_2 ; 1\, q | f_1\,m_1 \rangle,
\end{align}
and similarly for $\D^+_q\rightarrow(-1)^q\D^-_{-q}$. Here, $\eta_{1,2}$ accounts for other quantum numbers associated to the states considered, e.g.~$e$ and $g$.

In the main text, we encountered the scalar product
\begin{align}
	\Dv^+ \cdot \Dv^- =&\, \sum_{q,q'} (-1)^{q+q'} \mbf e_{-q}\cdot \mbf e_{-q'}^* \D^+_{q} \D^-_{q'} \nonumber\\
	=&\, \sum_{q} \D^+_{q} \D^-_{q}.
\end{align}
Since the scalar product of two vectors gives rise to a scalar under rotations, this shows that the sum $\sum_{q} \D^+_{q} \D^-_{q}$ is a scalar and thus commutes with the total angular momentum operator.

%%%%%%%%%%%%%						%%%%%%%%%%%%%
%%%%%%%%%%%%%		    APPENDIX		%%%%%%%%%%%%%
%%%%%%%%%%%%%						%%%%%%%%%%%%%

\section{Dark states from single-particle basis\label{app:darks_single}}

We prove here that the states of Eqs.~(\ref{eq:dark_n2square}) and (\ref{eq:dark_n2V}) are the only existing dark states for $n=2$ using the single-particle Fock basis.

In general, a dark state is a superposition of different Fock states that has to fulfill Eq.~(\ref{eq:dark_condition_qf}).
As argued in Sec.~\ref{ssec:interference}, we can assume that all Fock states in the superposition have the same number of excitations $n_e$ and magnetic number $M$.
Each of these Fock states has different possible decay channels and all need to be killed either by Pauli exclusion or through destructive interference with the decay of a different state.
Because $n=2$, there are at most two different Fock states which can decay to the same final state $\ket{\phi_f}$ via the same polarization $q$.
Therefore, Fock states can only interfere in pairs in this case.

When a Fock state has a decay channel $(\ket{\phi_f},q)$ that no other Fock state has, we call it a unique decay channel. Fock states with unique decay channels can not be part of a dark state superposition. This can be used to show the absence of dark states for various cases: states with two excitations, states with $M\neq0$ for the multi-$\square$ level structure, and any state for the multi-$\Lambda$ structure.
For double-excited states, it is straightforward to see that there exist no two states of the form $\ket{e_m\,e_n}$ that can decay to the same final state via the same polarization.
For the multi-$\square$ level structure, one can show that in any set of states with $M\neq0$ there is always one state which has a unique decay channel. The same argument applies to the multi-$\Lambda$ level structure for any $M$.

For the remaining cases, it follows from the previous considerations that we can write a dark state ansatz generically as
\begin{equation}
	\ket{D_M}_{\{f_g,f_e\}} = \sum_{m} \alpha^{(f_g,f_e,M)}_m \ket{g_{M-m}e_{m}},
\label{eq:dark_ansatz}
\end{equation}
where the sum is restricted to $-f_g\leq M-m \leq f_g$ and $-f_e\leq m \leq f_e$. The corresponding amplitudes $\alpha^{(f_g,f_e,M)}_m$ need to fulfill the system of equations (\ref{eq:dark_condition_qf}). Since the Clebsch-Gordan coefficients are real, the amplitudes can be assumed to be real too.

For Eq.~(\ref{eq:dark_ansatz}) to be dark, each decay channel $(\ket{\phi_f},q)$ has to be either blocked by Pauli exclusion, or cancelled out through interference between a pair of states. In the latter case, for a given final state $\ket{g_mg_{M-m}}$ and decay polarization $q$, the pair of interfering states is given by $\ket{g_me_{M+q-m}}$ and $\ket{e_{m+q}g_{M-m}}$. Both these states can decay to $\ket{g_mg_{M-m}}$ by emitting a photon with $q$ polarization. We call this pair of states a \emph{conjugate} pair with respect to $q$, or $q$-pair.

Interference happens between conjugate pairs of states.
For a conjugate pair in Eq.~(\ref{eq:dark_ansatz}) to interfere destructively, it must fulfill (dropping the superscript of the amplitudes for simplicity)
\begin{align}
	\D^-_q \left( \alpha_{M+q-m}\ket{g_me_{M+q-m}} + \alpha_{m+q}\ket{g_{M-m}e_{m+q}} \right) =0.
\end{align}
This implies that
\begin{equation}
	C^q_{M-m}\,\alpha_{M+q-m} = C^q_m\,\alpha_{m+q}.
\label{eq:pair_equation}
\end{equation}
Note that the indices have to fulfill $-f_g \leq m, M-m \leq f_g$, and $-f_e \leq m+q,M+q-m \leq f_e$.
Thus, every distinct $q$-pair in Eq.~(\ref{eq:dark_ansatz}) gives rise to an equation of the form (\ref{eq:pair_equation}), and the whole system of equations is made of intertwined pairs.

\begin{figure}[t!]
\centering
	\includegraphics[width=\columnwidth]{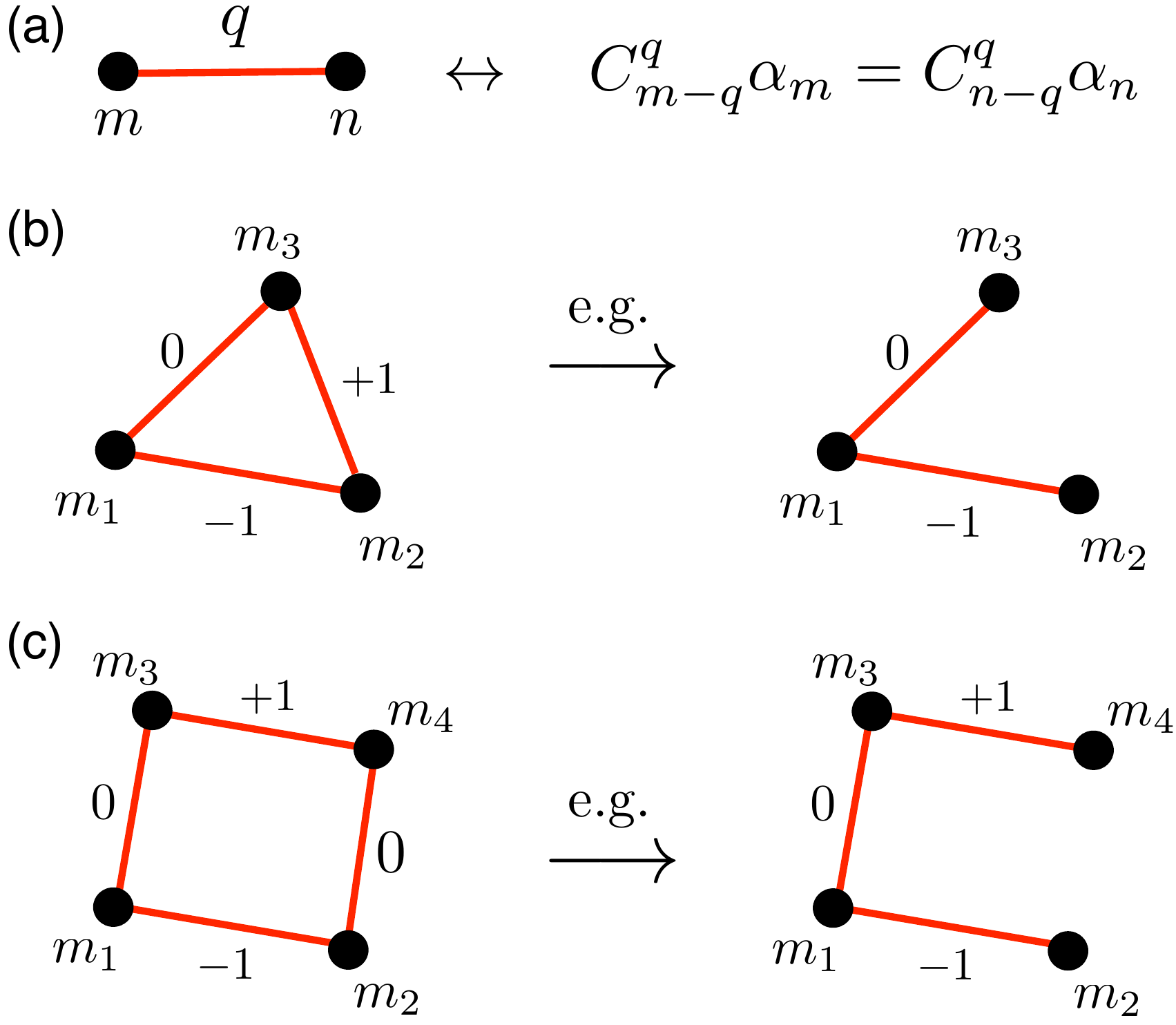}
	\caption{(a) Graphical representation of dark state equation for a conjugate pair of states $\ket{g_ne_m}$ and $\ket{g_me_n}$ for polarization $q$. (b) Reduction of triangle loop due to Eq.~(\ref{eq:cycling3}). (c) Reduction of square loop due to Eq.~(\ref{eq:cycling4}).}
	\label{fig:darknetworks}
\end{figure}

It is useful to graphically represent the system of equations by a network, see Fig.~\ref{fig:darknetworks}. Each node of the network represents one of the Fock states involved in the superposition (\ref{eq:dark_ansatz}) and is labelled by the index $m$ of its amplitude $\alpha_m$. Each line connecting two nodes has a polarization $q$ attached and represents an equation of the form (\ref{eq:pair_equation}) connecting the corresponding states of the nodes. 
The network corresponding to each ansatz (\ref{eq:dark_ansatz}) is connected, and the total number of lines is equal to the total number of equations to be solved. The network can in principle have more lines than nodes, such that the system of equations appears a priori indeterminate.

Due to the properties of Clebsch-Gordan coefficients, however, some equations are redundant and can be eliminated. Specifically, for the level structure $f\leftrightarrow f+1$ one can show that the Clebsch-Gordan coefficients fulfill the 3-state cycling property
\begin{equation}
	\frac{C^0_{m-1}}{C^0_{m+1}} \frac{C^-_{m+1}}{C^-_m} \frac{C^+_m}{C^+_{m-1}} = 1,
\label{eq:cycling3}
\end{equation}
where $|m\pm1|\leq f$, as well as the 4-state cycling property
\begin{equation}
	\frac{C^0_{m}}{C^0_{n}} \frac{C^-_{n+1}}{C^-_m} \frac{C^0_{m-1}}{C^0_{n+1}} \frac{C^+_n}{C^+_{m-1}} = 1,
\label{eq:cycling4}
\end{equation}
where $|m|,|m-1|\leq f$, and $|n|,|n+1|\leq f$. Note that the properties (\ref{eq:cycling3}) and (\ref{eq:cycling4}) can be used to derive higher-order cycling identities.
For the level structure $f\leftrightarrow f$, the property (\ref{eq:cycling3}) is not valid, but the property (\ref{eq:cycling4}) is fulfilled for $n=-m$, which is relevant for the $M=0$ case. 

Using Eqs.~(\ref{eq:cycling3}) and (\ref{eq:cycling4}), one can show that the $q$-pair equations involved in closed loops of the network are not linearly independent. Therefore, one of the equations forming the closed loop can be eliminated from the system of equations, as shown in Fig.~\ref{fig:darknetworks} for a triangle and a square loop. Following this procedure iteratively, one can reduce the number of lines in the network until the number of lines is $V-1$ for $V$ nodes. Once this is done, a unique solution to the system of equations represented by the network can be straighforwardly found.

This can be used to prove that the states (\ref{eq:dark_n2square}) and (\ref{eq:dark_n2V}) are the only dark states for $n=2$. For this, one can show first that for the multi-$V$ level structure, and for the multi-$\square$ level structure with $M=0$, all states in the dark state superposition ansatz of Eq.~(\ref{eq:dark_ansatz}) have a conjugate pair state to interfere with. In other words, for these cases the system of equations $\D^-_q\ket{D}=0$ is made of a set of $q$-pair equations such as Eq.~(\ref{eq:pair_equation}). The network associated to these equations can then be solved as explained above.

%%%%%%%%%%%%%						%%%%%%%%%%%%%
%%%%%%%%%%%%%		    APPENDIX		%%%%%%%%%%%%%
%%%%%%%%%%%%%						%%%%%%%%%%%%%

\section{Total angular momentum states for identical particles\label{app:totalFstates}}

To find the allowed total angular momentum states of $n\geq3$ identical fermions we employ the method described in Ref.~\cite{devanathan2006angular}. It consists of writing down all single-particle Fock states allowed by the Pauli exclusion principle, calculating their total angular momentum projection ($M=m_1+m_2+m_3+\ldots$), and then inferring from this the values of $F$ associated with those same values of $M$. For example, for $n=3$ fermions with angular momentum $f_g=3/2$ we have four allowed states: $\ket{g_{-3/2}\,g_{-1/2}\,g_{1/2}}$, $\ket{g_{-3/2}\,g_{-1/2}\,g_{3/2}}$, $\ket{g_{-3/2}\,g_{1/2}\,g_{3/2}}$, and $\ket{g_{-1/2}\,g_{1/2}\,g_{3/2}}$, with $M=-3/2,-1/2,1/2,3/2$, respectively. Therefore, the total $F$ must be $3/2$, since it leads to the same number of states (4) and same values of $M$ as in the single-particle basis.

When not all atoms are identical, we proceed in two steps. First, we find the allowed total angular momenta of the subset of atoms that are identical, i.e.~that are in the same orbital state $g$ or $e$. Then, we combine the result with the rest of the atoms using the usual rules of angular momentum addition. For example, to find out the allowed $gge$ states, we first construct the states $\ket{f_{12},M}_{gg}$, where $f_{12}$ can only be an even number between $0$ and $2f_g$. Each value of $f_{12}$ is then added to the $f_e$ of the remaining atom, similarly to Eq.~(\ref{eq:FM_statesn2_def}), to form $\ket{(f_{12})F,M}_{gge}$, where $F=|f_{12}-f_e|,\ldots,f_{12}+f_e$.

%%%%%%%%%%%%%						%%%%%%%%%%%%%
%%%%%%%%%%%%%		    APPENDIX		%%%%%%%%%%%%%
%%%%%%%%%%%%%						%%%%%%%%%%%%%

\section{Dark superposition of different \texorpdfstring{$f_{12}$}{f12} states\label{app:dark_superposition}}

We provide here an example of how to explicitly construct a dark state made of a superposition of different angular momentum states. Specifically, we consider $n=3$ atoms with $f_g=3/2\leftrightarrow f_e=3/2$. As argued in the main text (Sec.~\ref{ssec:darks_n3}), there exists a dark state superposition of the form
\begin{equation}
	\ket{D_M}^{n=3}_{\{3/2,3/2\}} = \alpha \ket{(0)3/2,M}_{gge} + \beta \ket{(2)3/2,M}_{gge}.
\end{equation}
Following the arguments of the main text, the values of $\alpha$ and $\beta$ depend only on reduced dipole matrix elements and are hence independent of $M$. Therefore, we consider $M=3/2$.

Following the rules to construct states of App.~\ref{app:totalFstates} one can show that
\begin{align}
	&\,\ket{(0)3/2,3/2}_{gge} \nonumber\\
	&\, \qquad = \frac{1}{\sqrt{2}} \left( \ket{g_{3/2}\,g_{-3/2}\,e_{3/2}} - \ket{g_{1/2}\,g_{-1/2}\,e_{3/2}} \right), \\
	&\,\ket{(2)3/2,3/2}_{gge} \nonumber\\
	&\, \qquad = \frac{1}{\sqrt{10}} \left( 2\ket{g_{-1/2}\,g_{3/2}\,e_{1/2}} - 2 \ket{g_{1/2}\,g_{3/2}\,e_{-1/2}} \right. \nonumber\\
	&\, \qquad\qquad \left. - \ket{g_{-3/2}\,g_{3/2}\,e_{3/2}} - \ket{g_{-1/2}\,g_{1/2}\,e_{3/2}} \right).
\end{align}
Since we showed that all $q=0,\pm1$ decay channels acting on states with the same $F$ are linearly dependent from each other, we only need to consider one single $q$.
Acting with $\D^-_0$ on $\ket{D_M}^{n=3}_{\{3/2,3/2\}}$ we find
\begin{align}
	\D^-_0 \ket{D_{3/2}}^{n=3}_{\{3/2,3/2\}} = \left( -\frac{3}{\sqrt{30}}\alpha + \frac{3}{5\sqrt{6}}\beta \right) \ket{g_{1/2}\,g_{-1/2}\,g_{3/2}}.
\end{align}
Thus, $\alpha=\beta/\sqrt{5}$. The normalized dark state is then given by
\begin{equation}
	\ket{D_M}^{n=3}_{\{3/2,3/2\}} = \frac{1}{\sqrt{6}} \left( \ket{(0)3/2,M}_{gge} + \sqrt{5}\, \ket{(2)3/2,M}_{gge} \right).
\end{equation}
It can be checked explicitly that this state indeed fulfills $\D^-_q\ket{D_M}^{n=3}_{\{3/2,3/2\}}=0$ for all $q$ and $M$.

\vfill

\bibliography{subrad_multilevel_pauli_bibliography}

\end{document}